\documentclass[12pt]{article}
\usepackage{amssymb}
\usepackage{amsmath}

\def\beq{\begin{equation}}\def\eeq{\end{equation}}
\def\bea{\begin{eqnarray}}\def\eea{\end{eqnarray}}
\begin{document}
\title{Bosonic Fields in Causal Set Theory}
\author{Roman Sverdlov
\\Mathematics Department, University of New Mexico}
\date{March 5, 2021}
\maketitle
\begin{abstract}
\noindent In this paper we will define a Lagrangian for scalar and gauge fields on causal sets, based on the selection of an Alexandrov set in which the variations of appropriate expressions in terms of either the scalar field or the gauge field holonomies around suitable loops take on the least value. For these fields, we will find that the values of the variations of these expressions define Lagrangians in covariant form.
\end{abstract}

\section{Introduction}

Causal set theory \cite{Nowhere, Surya} is a model of discrete spacetime where the only structure is a light cone causal relation, described by partial ordering, $\prec$. The relation $x \prec y$ is equivalent to the statement that $x$ is within the past light cone of $y$ or, equivalently, $y$ is within the future light cone of $x$. It has been shown that this partial ordering reproduces manifold structure up to conformal scaling \cite{Hawking, Malament}. In discrete setting, the conformal scaling is determined by the assumption that each point takes up the same volume and, therefore, causal structure becomes identical to geometry. At the same time, due to quantum fluctuations of gravity on the small scale, the manifold structure might break down. If one speculates that the well defined partial ordering continues to exist, this would provide a way of generalization Lagrangian densities into that setting. In general, that relation obeys the following axioms:

1) If $x \prec y$ and $y \prec z$ then $z \prec z$.

2) There is no point $x$ that satisfy $x \prec x$.

3) If $x \prec y$ then the number of points $z$ satisfying $x \prec z \prec y$ is at most finite

The axioms $1$ and $2$ tell us that $\prec$ is a partial ordering. An axiom 3 tells us that spacetime is discrete. We, furthermore, define $\preceq$ relation as follows:

{\bf Definition:} We say that $x \preceq y$ if either $x \prec y$ or $x =y$.

 In light of the discreteness, we have a \emph{direct neighbor} relation given as follows:

{\bf Definition:} We say that $x \prec^* y$ if the relation $x \prec y$ holds and one can not find $z$ satisfying $x \prec z \prec y$.

The approach of causal set project is to start from the situation where the manifold structure exists, and extract the quantities that can be generalized to the situations without a manifold structure. A good example is the way distances are defined \cite{Distance1, Distance2}. Consider the flat Minkowski space and suppose we are interested in a distance between two timelike separated points. We can select a coordinate system where both of these points are on $t$-axis. Since the length element is $\sqrt{dt^2 - \vert d \vec{x} \vert^2}$, the future-directed timelike curve that \emph{maximizes} the distance will be the segment of the $t$-axis itself. That is, a straight line. This can be generalized to the curved space: a timelike geodesic is the future directed curve that maximizes the distance. \emph{This, in turn, can be generalize to the causal set}. In particular, the discretized geodesic segment on a causal set is a chain of points $a \prec c_1 \prec \dots \prec c_{n-1} \prec b$ that is of maximal length, in a sense that if one finds some other set of points $a \prec d_1 \prec \cdots \prec d_{m-1} \prec b$, one always has $m \leq n$. Thus, geodesic distance is defined as
\begin{equation} \tau (a,b) = \max \{n \vert \exists c_1, \cdots, c_{n-1} (a \prec c_1 \prec \cdots \prec c_{n-1} \prec b) \} \label{Distance} \end{equation}
\emph{This equation is well defined without manifold structure}. Therefore, despite the fact that it was \emph{motivated} by the manifold, this definition carries over to non-manifoldlike settings.

The other ingredient in this project is to describe the otherwise-known fields and Lagrangians within this context. For example, outside of causal set theory, Lagrangians have derivative signs, which carry Lorentzian indices. Since these indices refer to coordinate axes that we no longer have, we can not use them. Instead, they have to be replaced with something that is solely based on causal relations alone.

Some proposals regarding Delambertians on the causal set have been made \cite{Delambertian1, Delambertian2}, but they encounter nonlocality issues \cite{Nonlocality}. The source of the non-locality is the fact that the Lorentzian neighborhood of any given point is the vicinity of its light cone, which stretches infinitely far coordinate-wise and has infinite volume. It is claimed, however, that the contributions coming from the ``tails" of the light cone cancel each other out, while the contributions coming from ``the middle" would add up, with the remainder being Lorentz invariant. While these results were confirmed numerically, on a theoretical level it is not as obvious as one might hope. For one thing, it turns out that different Delambertians are needed for different dimensions.

Another set of proposals \cite{paper1, paper3} was based on \emph{Alexandrov set} $\alpha (p,q) = \{r \vert p \preceq r \preceq q \}$ (see p.9 of \cite{Dynamics}) which, in case of Minkowski space, looks like a diamond (see \cite{Solodukhin}). We assume that Alexandrov set is large enough to contain large number of points yet small enough for us to be able to assume that the fields behave linearly within that set. We then search for the linear combination of various integrals over the fields over that region that would result in the Lagrangians that we are looking for. Due to the fact that this region is diamond-shape, the integrals end up non-trivial. Consequently, the coefficients that would give us the result that we need are non-trivial as well. And these non-trivial integrals, as well as the coefficients, end up being dimension-dependent.

In this paper we propose to treat scalar field in such a way that is both local and dimensionally independent. By dimensional independence we mean that we \emph{do} allow for the \emph{overall} factor to have dimensional dependence (such is the case, for example, in Eq \ref{MultipletLagrangian}) but we would like to avoid having a separate dimensional dependence in the coefficient next to one of the terms. Our approach is as follows. We first determine the direction of its gradient, $v^{\mu} = \partial^{\mu} \phi / \vert \partial^{\mu} \phi \vert$, and then determine its Lagrangian as $(v^{\mu} \partial_{\mu} \phi)^2$. Just like in papers \cite{paper1, paper3}, however, it is assumed that the field behaves linearly. The advantage of this approach is that some theories of gravity predict the changes of dimensions on small scale \cite{2Dim1, 2Dim2}. While these specific models might not be compatible with the causal set proposals described in this paper, the current paper might accommodate other similar situations where the dimensions would change. 

The other goal of the current paper is to extend the discussion from scalar fields on the causal set to the electromagnetic fields. The only other proposals of electromagnetic fields that we are aware of are the unpublished work by the author of the current paper \cite{GaugeFields, EMedges}. As far as we are aware, there are no proposals of electromagnetic Lagrangians within wider causal set community. It should be acknowledged though that the current proposal utilizes the notions of ``electric" and ``magnetic" field that $F_{\mu \nu}$ would only reduce to in $d=3+1$ case. At the same time, however, we do not have any non-trivial coefficients that are the functions of $d$. Therefore, this approach carries over to other dimensions, although it would lose its physical meaning.

\section{Spin 0 fields} \label{Scalar}

\subsection{Few words about discrete structures} \label{FewWords}

Before we proceed with the main topic of this chapter, let us clarify some points regarding the discrete structures that the reader might have. When one first encounters a causal set, one might naturally ask: what exact structure does it have? The answer will never be too specific. Essentially, any partially ordered set can be thought of as a causal set. For example, some toy models have been proposed of dynamically generating a causal set, which are known as \emph{dynamical percolation} models \cite{Percolation} and the resulting structures do not resemble a manifold, no matter what scale one looks at. This makes them irrelevant to the kind of structures we might be interested in. One possible way of making a causal set appear like a manifold is to use a cubic lattice. However, it breaks down Lorentz invariance in a rather obvious way. The other approach is to utilize Robb's description of axiomatizing Minkowski space \cite{Robb1, Robb2} and then adapt it to discretization and curvature. But, even in the continuum flat Minkowskian setting that list of axioms is rather complicated, so it might get even more complicated if one adds discretization and curvature into the mix. There are other, ``softer" approaches to the issue. In particular, while they don't provide exclusive criteria for a causal set being manifoldlike, they provide some statistical hallmarks of this (see, for example, \cite{ManifoldlikeThesis}). But since the exclusive definition of manifoldlike causal set is not given, this is not sufficient. The other approach is to simply use Poisson process (see \cite{DiscreteGravity}, pp 9-11). But the fact that it relies on already existing manifold seems to contradict the idea that the scattered points are the building blocks of spacetime itself, which is the spirit of discrete theories. Furthermore, the manifold on which Poisson process is performed is assumed to be continuous on all scales, in contradiction to quantum fluctuations of gravity that are predicted to occur on Planck scale. And finally, in light of the fact that Lorentzian $\epsilon$-neighborhood is nonlocal (it fills the vicinity of a light cone) every point will have infinitely many direct neighbors, most of which are arbitrarily far away coordinate-wise, which poses a big problem \cite{Nonlocality}. There were some proposals on how to curb it, made by the author of the current paper (see, for example, Section 2.2 of \cite{ContinuousMeasurement}, pages 6,7 of \cite{EMedges}, Sections 2.2 and 2.3 of \cite{PhaseSpacetime} and most of \cite{Proceedings}). However, those proposals are based on altering the ontology of spacetime itself in one way or the other and, therefore, they are too controversial for the current paper.

To make long story short, the exact structure of a causal set is one of the unanswered questions in causal set theory that deserves a separate paper. As far as this paper is concerned, its focus is (non-gravitational) quantum field theory on a causal set as opposed to its geometry. Therefore we will simply not attempt to speculate as to which of these structure(s) we are working with. Instead, we will simply \emph{assume} that, one way or the other, causal set approximates a manifold starting from certain scales and, one way or the other, the non-localities have been curbed. The way we would make the specifics of small-scale structures irrelevant is by relying on statistical properties of a causal set that one would expect to see on the scales where the manifold structure emerges. The way we would do that is that, instead of defining Lagrangian density based on the direct neighbors of any given point, we will define it in terms of  Alexandrov sets that contain large number of points, so that statistical approximations can be reached. The density of points will be assumed to be large enough for those neighborhoods to still be regarded as very small, despite having a very large number of points.

We will enforce this idea in the following way. Instead of using direct neighbors in our definition of Lagrangian density, we will be using Alexandrov set, $\alpha (p,q)$. We will then introduce \emph{two} fundamental lengths. One, which is denoted by $\tau_1$, is the statistical average distance between two neighboring points, and the other, which we will denote by $\tau_2$, is the distance between points $p$ and $q$ if Alexandrov set $\alpha (p,q)$ is to be used for the definition of Lagrangian density. If we make $\frac{\tau_2}{\tau_1}$ very large, we would have enough points for a very good approximation to the statistical limit to be reached within $\alpha (p,q)$ and then the specifics of the discrete structure would not be relevant. At the same time, if we also make sure that both $\tau_1$ and $\tau_2$ are very small in our scale, we would be able to use linear approximation in that region, like we do througout this paper. In order for quantum fluctuations of gravity not to be felt on the scale $\tau_2$, we would have to claim that Planck scale is much smaller. For example, we could claim that Planck scale is $\tau_1$. Although it is also possible that Planck scale is somewhere in between $\tau_1$ and $\tau_2$. We will agree that Planck scale is some function of $\tau_1$ and $\tau_2$, but we will leave the speculation of what it might be for future research.

It is important to emphasize that this relies on the following physical assumption. Namely, we are assuming that the non-gravitational Lagrangian densities are not local but semi-local, where semi-locality is defined based on the scale of $\tau_2$. On the local scale, defined in terms of $\tau_1$, we have purely gravitational processes or lack thereof (we are leaving this for other papers). But, on the scale of $\tau_2$, we have the non-gravitational Lagrangian densities. And, due to the \emph{physical} assumption that non-gravitational Lagrangian densities, indeed, act on the scale of $\tau_2$, we are able to ignore the small scale behavior of the causal set when we will be modeling those. The other assumption that is being made is that the small scale behavior does not create any anomalies that would throw away some of the implicit assumptions we are making in our geometric analysis. This is, indeed, a big assumption to make. Nevertheless, for the purposes of this paper, we will work under this assumption. 

\subsection{Real scalar field} \label{RealScalar}

Let us now proceed to defining quantum field theory on a causal set $\mathcal C$ and a real scalar field $\phi \colon {\mathcal C} \mapsto \mathbb{R}$. We define Lagrangian density $\mathcal L$ in the following way. We will break it into two parts: one part comes from timelike separated points, ${\mathcal L}_t$, and the other part comes from spacelike separated points, ${\mathcal L}_s$. And then the actual Lagrangian will be the difference between the two:
\begin{equation} {\mathcal L}_t (\phi; x) =\frac{1}{\tau_2^2} \min \big\{( \phi (q) - \phi (p) )^2 \big\vert  p \prec x \prec q; \tau (p,q) = \tau_2 \big\} \label{Lt} \end{equation}
\begin{equation}{\mathcal L}_s (\phi; x) = \frac{1}{\tau_2^2} \min \Big\{ \max \big\{ ( \phi (r) - \phi (s) )^2 \big\vert p \prec^* r \prec^* q \; , \; p \prec^* s \prec^* q \big\} \Big\vert \tau(p,q) = \tau_2 \Big\} \label{Ls} \end{equation}
\begin{equation} {\mathcal L} (\phi; x) = {\mathcal L}_t (\phi; x) - {\mathcal L}_s (\phi; x) \label{ScalarSubtraction} \end{equation}
where $\tau_2$ is treated as a constant, which was discussed in the last two paragraphs of Section \ref{FewWords}. Since the construction is invariant, we claim that ${\mathcal L}_t$ and ${\mathcal L}_s$, on their own, would produce Lorentz invariant expressions. The only reason we need to subtract them is that those Lorentz invariant expressions will be multiplied by step functions (see Eq \ref{LtScalar} and Eq \ref{LsScalar}) and we need the subtraction in order to cancel those step functions. However, these step functions are taken over Lorentz invariant expressions and, therefore, they are Lorentz invariant. Therefore, Lorentz invariance is not the issue; the issue is that the Lorentz invariant expressions that we got happened not to be the ones we are looking for; and that's why we subtract one of them from the other in order to fix it. In any case, let us go ahead and derive Eq \ref{LtScalar} and \ref{LsScalar}. We will do it by cases.

{\bf Case 1:} The gradient of $\phi$ is timelike. That is, $\partial^{\mu} \phi \partial_{\mu} \phi >0$.

For paedagogical purposes, let us derive Lagrangian in two different ways. One way is to agree upon a coordinate system, which will be simpler, and the other way is to to it for arbitrary coordinates, which would be more complicated.

{\bf Case 1: Conveniently selected coordinate system} Lets start with the simpler way. Consider the coordinate system in which $t$-axis is parallel to the gradient of $\phi$; that is,
\begin{equation} \partial_{\mu} \phi = \delta^0_{\mu} \partial_0 \phi \end{equation}
As always, we assume $\phi$ is linear and, therefore, its gradient is a constant. In this case, in order to minimize $(\phi (q) - \phi (p))^2$, we simply have to minimize $(q^0 - p^0)^2$. Since our constraint is $\tau (p,q) \approx \tau_2$, this corresponds to the situation when
\begin{equation} q^{\mu} - p^{\mu} \approx \tau_2 \delta^{\mu}_0 \end{equation}
and, therefore,
\begin{equation} \phi (q) - \phi (p) \approx (q^{\mu} - p^{\mu}) \partial_{\mu} \phi \approx \tau_2 \delta^{\mu}_0 \partial_{\mu} \phi = \tau_2 \partial_0 \phi = \tau_2 \sqrt{\partial^0 \phi \partial_0 \phi } = \tau_2 \sqrt{\partial^{\mu} \phi \partial_{\mu} \phi} \end{equation}
Consequently,
\begin{equation} {\mathcal L}_t  (\phi, x) \approx \frac{1}{\tau_2^2} \big( \tau_2 \sqrt{\partial^{\mu} \phi \partial_{\mu} \phi} \big)^2 = \partial^{\mu} \phi \partial_{\mu} \phi \end{equation}
Let us now look at ${\mathcal L}_s$. Once again, we are in a coordinate system where the gradient of $\phi$ is parallel to $t$-axis,
\begin{equation} \partial_{\mu} \phi = \delta^0_{\mu} \partial_0 \phi \end{equation}
Therefore, in order to minimize $(\phi (r) - \phi (s))^2$, we need to minimize $(r^0 - s^0)^2$. Now, geometrically, the conditions that $p \prec^* r \prec^* q$ and $p \prec^* s \prec^*q$ imply that $r$ and $s$ are on the equator of Alexandrov set (where by ``equator" we mean the sphere formed by intersection of two light cones). If the axis of Alexandrov set is parallel to $t$-axis, then the equator of the Alexandrov set will be perpendicular to $t$-axis and, therefore, we would have $r^0 - s^0 =0$ which would imply that $(\phi (r) - \phi (s))^2 \approx 0$. Since, in this sub-section, we are assuming the scalar field is real, the square can't possibly be negative. Thus, we conclude that $0$ corresponds to its minimum. Since ${\mathcal L}_s$ is defined in terms of the minimum, we conclude that
\begin{equation}{\mathcal L}_s (\phi, x) \approx 0 \end{equation}

{\bf Case 1: General coordinate system} Let us now prove the same thing without alligning $t$-axis in any particular way. Let us start by evaluating ${\mathcal L}_t$. We assume that $\phi$ is approximately linear throughout the region of our concern. Therefore,
\begin{equation} \phi (q) - \phi (p) \approx (q^{\mu} - p^{\mu}) \partial_{\mu} \phi \end{equation}
Since both $q-p$ and $\partial \phi$ are timelike, the reverse Schwartz inequality applies. Thus,
\begin{equation} \vert \phi (q) - \phi (p) \vert \geq \vert q-p \vert \vert \partial \phi  \vert \approx \tau_2 \sqrt{\partial^{\mu} \phi \partial_{\mu} \phi} \end{equation}
We also know that $\geq$ sign is being replaced by $=$ sign when $q-p$ and $\partial^{\mu} \phi$ are parallel. Therefore, the mininum is
\begin{equation} \min \Big\{ \vert \phi (q) - \phi (p) \vert \Big\vert x \in \alpha (p,q) \; , \; \tau (p,q) = \tau_2 \Big\} \approx  \tau_2 \sqrt{\partial^{\mu} \phi \partial_{\mu} \phi} \end{equation}
Substituting this into Eq \ref{Lt}, we obtain
\begin{equation} {\mathcal L}_t \approx \frac{1}{\tau_2^2} (\tau_2 \sqrt{\partial^{\mu} \phi \partial_{\mu} \phi})^2 = \partial^{\mu} \phi \partial_{\mu} \phi \end{equation}
Let us now look at ${\mathcal L}_s$. I claim that ${\mathcal L}_s \approx 0$. Again, since the square can't possibly be less than $0$, in order to prove that the minimum is $0$ it is sufficient to show at least one situation when one gets $0$, and that situation would become a minimum by default. We claim that the scenario when this is approximately $0$ occurs, again, when $\partial \phi$ and $q-p$ are parallel. Recall from the Eq \ref{Ls} that we are interested in the situation where
\begin{equation} p \prec^*r \prec^*q \end{equation}
\begin{equation} p \prec^*s \prec^* q \end{equation}
Geometrically, this means that $r$ and $s$ are on the equator of Alexandrov set while $p$ and $q$ are on the poles. The coordinate separation between points at the poles and points at the equator is large, while the Lorentzian distance is small. In such a scenario, the following calculation holds:
\begin{equation} a \prec^* b \Rightarrow (a^{\mu} - b^{\mu})(a_{\mu} - b_{\mu}) \approx 0 \Rightarrow a^{\mu} a_{\mu} - 2a^{\mu}b_{\mu} + b^{\mu} b_{\mu} \approx 0 \Rightarrow a^{\mu}b_{\mu} \approx \frac{a^{\mu}a_{\mu} + b^{\mu}b_{\mu}}{2} \end{equation}
Therefore,
\begin{equation} p \prec^* r \Longrightarrow p^{\mu} r_{\mu} \approx \frac{p^{\mu} p_{\mu} + r^{\mu}r_{\mu}}{2} \end{equation}
\begin{equation} p \prec^* s \Longrightarrow p^{\mu} s_{\mu} \approx \frac{p^{\mu} p_{\mu} + s^{\mu}s_{\mu}}{2} \end{equation}
\begin{equation} r \prec^* q \Longrightarrow q^{\mu} r_{\mu} \approx \frac{q^{\mu} q_{\mu} + r^{\mu}r_{\mu}}{2} \end{equation}
\begin{equation} s \prec^* q \Longrightarrow q^{\mu} s_{\mu} \approx \frac{q^{\mu} q_{\mu} + s^{\mu}s_{\mu}}{2} \end{equation}
We can then perform the following calculation:
\begin{equation} (q^{\mu} - p^{\mu})(r_{\mu} - s_{\mu}) = q^{\mu} r_{\mu} - q^{\mu} s_{\mu} - p^{\mu} r_{\mu} + p^{\mu}s_{\mu} \approx \nonumber \end{equation}
\begin{equation} \approx  \frac{q^{\mu} q_{\mu} + r^{\mu}r_{\mu}}{2} - \frac{q^{\mu} q_{\mu} + s^{\mu}s_{\mu}}{2} -  \frac{p^{\mu} p_{\mu} + r^{\mu}r_{\mu}}{2}  +  \frac{p^{\mu} p_{\mu} + s^{\mu}s_{\mu}}{2} = 0 \end{equation}
Since the square can't be less than $0$, and we just found $0$, the $0$ has to be a minimum. Thus, we conclude that
\begin{equation} {\mathcal L}_s (\phi; x) \approx 0 \end{equation}

{\bf Case 2:} Now suppose that $\partial_{\mu} \phi$ is spacelike. If we try to do it in general coordinate system, it would be more complicated than in Case 1, although it can be done. On the other hand, the method of selecting coordinate system where $x^1$-axis points along the gradient of $\phi$ is equally valid and, at the same time much simpler. So let us just stick to this method, and leave more general method as an exercize for the reader.

If we allign $x^1$-axis with the direction of $\partial_{\mu} \phi$, then the maximal value of $(\phi (r) - \phi (s))^2$ is proportional to the maximal value of $(r^1 - s^1)^2$. So our goal is to minimize the maximum of $(r^1 - s^1)^2$. Note that if we perform Lorentz transformation that mixes $(x^0, x^2, \cdots, x^{d-1})$, this does not affect $x^1$, and, therefore, does not change the maximum of $(r^1 - s^1)^2$. At the same time, such transformation can simplify things a great deal by making sure that
\begin{equation} q^2 -p^2 = \cdots = q^{d-1} -p^{d-1} = 0 \end{equation}
Now, the intersection of Alexandrov set with $x^0x^1$-plane will produce points $u$ and $v$ that satisfy the following equations:
\begin{equation} p^1 - u^1 = u^0 -p^0 \label{1} \end{equation}
\begin{equation} v^1 -p^1 = v^0 - p^0 \label{2} \end{equation}
\begin{equation} q^1 -u^1 = q^0 - u^0 \label{3} \end{equation}
\begin{equation} v^1 -q^1 = q^0-v^0 \label{4} \end{equation}
By adding Eq \ref{1} and Eq \ref{3} we obtain
\begin{equation} p^1 + q^1 -2u^1 = q^0 -p^0 \end{equation}
which implies that
\begin{equation} u^1 = \frac{p^1 +q^1}{2} - \frac{q^0-p^0}{2} \label{r1} \end{equation}
On the other hand, by adding Eq \ref{2} and Eq \ref{4} we obtain
\begin{equation} 2v^1-p^1-q^1=q^0-p^0 \end{equation}
which implies that
\begin{equation} v^1 = \frac{p^1+q^1}{2} + \frac{q^0-p^0}{2} \label{s1} \end{equation}
By subtracting Eq \ref{r1} from Eq \ref{s1} we obtain
\begin{equation} v^1-u^1 = q^0 -p^0 \end{equation}
Now, if $q^0 -p^0 > \tau_2$ this implies that $v^1 -u^1 > \tau_2$ and, therefore, $\max \{r^1 - s^1 \vert p \prec^* r \prec^* q; p \prec^* s \prec^* q \} > \tau_2$. On the other hand, if $q^0 - p^0 = \tau_2$, then $v^1 -u^1 = \tau_2$. Furthermore,$q^0 - p^0 = \tau_2$ implies $q^k - p^k =0$ and, therefore, the equator is just a ball of radius $\frac{\tau_2}{2}$, which means that $\tau_2$ is, indeed, the maximum of $r^1 -s^1$. Since for $q^0 - p^0 = \tau_2$ the maximum is $\tau_2$ and in all other cases the maximum is greater than $\tau_2$, then we conclude that the minimum of the maximum is $\tau_2$:
\begin{equation} \min \{ \max \{ r^1 -s^1 \vert p \prec^* r \prec^* q; p \prec^* s \prec^*q \} \vert p \prec x \prec q; \tau (p,q) = \tau_2 \} = \tau_2 \end{equation}
and, consequently,
\begin{equation} {\mathcal L}_s = \frac{1}{\tau_2^2} (\tau_2 \partial_1 \phi)^2 = \partial_1 \phi^2 = - \partial^{\mu} \phi \partial_{\mu} \phi \end{equation}
On the other hand, if we make any choice of $p$ and $q$ such that $q^1=p^1$, then we have $(\phi (q) - \phi (p))^2 \approx 0$. Since the square can't be negative, we conclude that $0$ must be its minimum and, therefore
\begin{equation} {\mathcal L}_s (\phi; x) \approx 0 \end{equation}
We can summarize the above results as follows. We introduce step function
\begin{equation} \sigma (\xi) =
\begin{cases}
0 , \xi< 0 \\
1, \xi \geq 0
\end{cases} \end{equation}
and then we obtain the following expressions:
\begin{equation} {\mathcal L}_t (\phi; x) \approx \partial^{\mu} \phi \partial_{\mu} \phi \; \sigma (\partial^{\mu} \phi \partial_{\mu} \phi) \label{LtScalar} \end{equation}
\begin{equation} {\mathcal L}_s (\phi; x) \approx \partial^{\mu} \phi \partial_{\mu} \phi (\sigma (\partial^{\mu} \phi \partial_{\mu} \phi)-1) \label{LsScalar} \end{equation}
which leads to
\begin{equation} {\mathcal L}_t (\phi; x) - {\mathcal L}_s (\phi; x) = \partial^{\mu} \phi \partial_{\mu} \phi \end{equation}
Thus, we define Lagrangian as this difference:
\begin{equation} {\mathcal L}_{\mathbb{R}} (\phi; x) = \frac{1}{\tau_2^2} \min \big\{( \phi (q) - \phi (p) )^2 \big\vert p \prec x \prec q; \tau (p,q) = \tau_2 \big\} - \nonumber  \end{equation}
\begin{equation} -  \frac{1}{\tau_2^2} \min \Big\{ \max \big\{ ( \phi (r) - \phi (s) )^2 \big\vert p \prec^* r \prec^* q \; , \; p \prec^* s \prec^* q \big\} \Big\vert \tau(p,q) = \tau_2 \Big\} \label{RealLagrangian} \end{equation}

\subsection{Complex scalar field} \label{ComplexScalar}

Let us extend our discussion from real scalar field to complex scalar field. One might be tempted to simply express complex scalar field as $\phi = \phi_1 +i \phi_2$, where $\phi_1$ and $\phi_2$ are both real, write $\partial^{\mu} \phi^* \partial_{\mu} \phi = \partial^{\mu} \phi_1 \partial_{\mu} \phi_1 + \partial^{\mu} \phi_2 \partial_{\mu} \phi_2$ and then use Section \ref{RealScalar} to model each term. The problem with this approach is that taking the maxima and minima of either real or imaginary part would not respect $U(1)$ symmetry.  To counter this objection, one might argue that if we successfully reproduce each term, we would successfully reproduce the sum, and the sum would respect $U(1)$ symmetry by default. However, there is a gap in that argument: it is not true that each term would be reproduced \emph{exactly}; it will only be \emph{approximated}. Therefore, \emph{up to that order of approximation} the $U(1)$ symmetry would be respected, but the error between continuum and discrete expression will not respect that symmetry. One might argue that since we assume that the above-mentioned error is too small to be detected (which is the assumption we have to make anyway) then the above violation of $U(1)$ symmetry would not be detectable either. So if we stick to the point of view that our belief in $U(1)$ symmetry is empirical in nature, we would be able to accept its small violations. However, for philosophical reasons, we would like to view $U(1)$ as fundamental rather than empirical, if at all possible. So let us see if we can find a model that would allow us to do that. To make it clear: we would still have error between discretized result and the continuum result, and we are not attempting to make it smaller than it otherwise would be. The only thing we are trying to do is to make sure that the error respects $U(1)$ symmetry.

One attempt that comes to mind is to look at $\tilde{\phi} (x, \theta) = Re (e^{i \theta} \phi)$ and then find $\theta (x)$ that would maximize $(\partial^{\mu} \tilde{\phi} \partial_{\mu} \tilde{\phi})\vert_{\theta(x)}$ at every point $x$, where $\partial^{\mu}$ and $\partial_{\mu}$ represent partial derivatives with respect to $x$, at fixed $\theta$.  While this could have worked in $0+1$ dimension, this would not work when the number of dimensions is higher than that. The issue is that it would produce $\vert \partial^{\mu} \phi \partial_{\mu} \phi \vert$ instead of $\partial^{\mu} \phi^* \partial_{\mu} \phi$. While in $1$ dimension it is true that $\vert \frac{\partial \phi}{\partial t} \frac{\partial \phi}{\partial t} \vert= \frac{\partial \phi^*}{\partial t} \frac{\partial \phi}{\partial t}$, in many dimensions it is no longer true: $\vert \partial^{\mu} \phi \partial_{\mu} \phi \vert \neq \partial^{\mu} \phi^* \partial_{\mu} \phi$. As a counter-example, consider $\phi = 2t - ix$. Then $\vert \partial^{\mu} \phi \partial_{\mu} \phi \vert= \vert 4-1 \vert =3$ while $\partial^{\mu} \phi^* \partial_{\mu} \phi = 4+1 =5$. Even if one were to use integration by parts to replace $\partial^{\mu} \phi^* \partial_{\mu} \phi$ with $\phi^* \Delta \phi$, we would still have $\vert \Phi \Delta \Phi \vert \neq \Phi^* \Delta \Phi$. So this approach wouldn't work either way.

The approach that \emph{would} work is to integrate over all possible choices of $\theta$. In this case, we have
\begin{equation} \int_0^{2 \pi} \partial^{\mu} Re (e^{i \theta} \phi) \partial_{\mu} Re (e^{i \theta} \phi) d \theta = \int_0^{2 \pi} \partial^{\mu} (\phi_1 \cos \theta - \phi_2 \sin \theta) \partial_{\mu} (\phi_1 \cos \theta - \phi_2 \sin \theta) d \theta = \nonumber \end{equation}
\begin{equation} = \int_0^{2 \pi} (\partial^{\mu} \phi_1 \partial_{\mu} \phi_1 \cos^2 \theta + \partial^{\mu} \phi_1 \partial_{\mu} \phi_2 \cos \theta \sin \theta + \partial^{\mu} \phi_2 \partial_{\mu} \phi_1 \sin \theta \cos \theta + \partial^{\mu} \phi_2 \partial_{\mu} \phi_2 \cos^2 \theta) = \nonumber \end{equation}
\begin{equation} = \pi \partial^{\mu} \phi_1 \partial_{\mu} \phi_1 + \pi \partial^{\mu} \phi_2 \partial_{\mu} \phi_2 = \pi \partial^{\mu} \phi^* \partial_{\mu} \phi \end{equation}
We then substitute $Re (e^{i \theta} \phi)$ into Sec \ref{RealScalar} to obtain
\begin{equation} \partial^{\mu} Re (e^{i \theta} \phi) \partial_{\mu} Re (e^{i \theta} \phi) \approx {\mathcal  L}_t (Re (e^{i \theta} \phi)) - {\mathcal L}_s (Re (e^{i \theta} \phi)) \end{equation}
which will lead us to define the Lagrangian as follows:
\begin{equation} {\mathcal L}_{\mathbb{C}} (\phi; x) = \frac{1}{\pi} \int_0^{2 \pi} {\mathcal L}_{\mathbb{R}} (Re (e^{i \theta} \phi)) d \theta \end{equation}
By substituting Eq \ref{RealLagrangian} we obtain
\begin{equation} {\mathcal L}_{\mathbb{C}} (\phi; x) = \label{ComplexLagrangian}\end{equation}
\begin{equation} =  \frac{1}{\pi \tau_2^2} \int_0^{2 \pi} \bigg(\min \big\{(Re(e^{i \theta} \phi (q))-Re(e^{i \theta} \phi (p)) )^2 \big\vert p \prec x \prec q; \tau (p,q) = \tau_2 \big\} - \nonumber  \end{equation}
\begin{equation} -  \min \Big\{ \max \big\{ ( Re (e^{i \theta} \phi (r)) - Re (e^{i \theta} \phi (s)) )^2 \big\vert p \prec^* r \prec^* q \; , \; p \prec^* s \prec^* q \big\} \Big\vert \tau(p,q) = \tau_2 \Big\} \bigg)  d \theta  \nonumber \end{equation}
Note that the integral over $\theta$ has not been discretized. This is because the only thing we are discretizing is spacetime manifold. Since $\theta$ does not live in a spacetime manifold, but instead it lives in $[0,2 \pi]$, we do not have to discretize it. On the other hand, the set of points over which maxima and minima are taken \emph{is} discretized.

\subsection{Complex spin-0 multiplet} \label{ComplexMultiplet}

Let us now generalize the results of Sec \ref{ComplexScalar} to multiplets. Suppose we have $n$-multiplet, which we would like to be symmetric under the symmetry group $G \supseteq SU(n)$. We will denote that multiplet by $\phi$ and we will denote its $i$-th component by $\phi_i$. Let $\pi_i$ be the projection of $\phi$ unto its $i$-th component, that leaves $i$-th component unchanged and sends the rest of the components to $0$:
\begin{equation} (\pi_i \phi)_j = \delta^j_i \phi_j \end{equation}
One can define Haar measure $\mu$ on $G$ (see \cite{Haar1, Haar2}) which would lead to the definition of its volume,
\begin{equation} \mu(G) = \int_G d \mu \end{equation}
 In particular, Eq 5.13 of \cite{Haar2} tells us
\begin{equation} \mu (SU(n)) = \sqrt{2^{n-1}n} \pi^{(n-1)(n+2)/2} \prod_{k=1}^{n-1} \frac{1}{k!} \end{equation}
and Eq 5.16 of \cite{Haar2} tells us
\begin{equation} \mu (U(n)) = \sqrt{2^{n+1}n} \pi^{n+1 \choose 2} \prod_{k=1}^{n-1} \frac{1}{k!} \end{equation}
 We can then perform the following calculation:
\begin{equation} \partial^{\mu} \phi^{\dagger} \partial_{\mu} \phi = \frac{1}{\mu (G)} \int_G \partial^{\mu} (U \phi)^{\dagger} \partial_{\mu} (U \phi) d \mu (U) = \frac{1}{\mu (G)} \int_G \sum_{i=1}^n \partial^{\mu} (\pi_i U \phi)^{\dagger} \partial_{\mu} (\pi_i U \phi) d \mu (U) = \nonumber \end{equation}
\begin{equation} =  \frac{1}{\mu (G)} \int_G \sum_{i=1}^n \partial^{\mu} (\pi_1 U \phi)^{\dagger} \partial_{\mu} (\pi_1 U \phi) = \frac{1}{\mu (G)}\partial^{\mu} (\pi_1 U \phi)^{\dagger} \partial_{\mu} (\pi_1 U \phi)  \sum_{i=1}^n 1 = \nonumber \end{equation}
\begin{equation} = \frac{n}{\mu (G)}\partial^{\mu} (\pi_1 U \phi)^{\dagger} \partial_{\mu} (\pi_1 U \phi) \end{equation}
where on the third equal sign we used the symmetry, which was produced as a consequence of integration over $G$, in order to replace $i$ with $1$. Therefore, we propose the Lagrangian to be
\begin{equation} {\mathcal L} = \frac{n}{\pi \mu (G)} \int_0^{2 \pi} ({\mathcal  L}_{\mathbb{R}} (Re (e^{i \theta} \pi_1 U \phi))d \theta  \end{equation}
If we now substitute Eq \ref{RealLagrangian} into this we obtain
\begin{equation}  {\mathcal L} (\phi; x) =  \nonumber \end{equation}
\begin{equation} =  \frac{n}{\pi \tau_2^2 \mu (G)} \int_0^{2 \pi} \bigg(  \min \big\{( Re(e^{i \theta} \pi_1 U \phi (q)) - Re(e^{i \theta} \pi_1 U \phi (p)) )^2 \big\vert p \prec x \prec q; \tau (p,q) = \tau_2 \big\} - \nonumber  \end{equation}
\begin{equation} - \min \Big\{ \max \big\{ ( ( Re(e^{i \theta} \pi_1 U \phi (r)) - ( Re(e^{i \theta} \pi_1 U \phi (s))  )^2 \big\vert p \prec^* r ,\prec^* q \; , \; p \prec^* s \prec^* q \big\} \Big\vert \nonumber \end{equation}
\begin{equation} \Big\vert  p \prec x \prec q, \tau(p,q) = \tau_2 \Big\} \bigg)  \label{MultipletLagrangian} \end{equation}

\section{Gauge Fields}

\subsection{Coupling between spin 0 fields and gauge fields} \label{Coupling}

Let us now discuss the gauge fields. We will start from electromagnetic field. In continuum setting with a coordinate system, we define a two-point function
\begin{equation} a(r, s) = \int_{\gamma (r,s)} A_{\mu} dx^{\mu} \end{equation}
where $\gamma (r,s)$ is a geodesic connecting points $r$ and $s$. In discrete setting, without coordinate system, we no longer have $A_{\mu}$ (since we can not define the Lorentz index) but we will still postulate the existence of two-point function $a(r,s)$ (since scalar quantities will still be well defined). Thus, $a(r,s)$ will replace $A_{\mu}$ as a definition of electromagnetic field.  In order to write the Lagrangian, we simply replace $\phi(z)$ with $e^{i a(x, z)} \phi (z)$ in Eq \ref{ComplexScalar}. Thus, we obtain  
\begin{equation} {\mathcal L}_{\mathbb{C}} (a,\phi; x) = \nonumber  \end{equation}
\begin{equation} = \frac{1}{\pi \tau_2^2} \int_0^{2 \pi} \bigg(\min \big\{(Re(e^{i \theta} e^{ia (x,q)} \phi (q))-Re(e^{i \theta} e^{i a (x,p)} \phi (p)) )^2 \big\vert p \prec x \prec q; \tau (p,q) = \tau_2 \big\} - \nonumber  \end{equation}
\begin{equation} -  \min \Big\{ \max \big\{ ( Re (e^{i \theta} e^{i a (x,r)} \phi (r)) - Re (e^{i a (x,s)} e^{i \theta} \phi (s)) )^2 \big\vert \nonumber \end{equation}
\begin{equation} \big\vert p \prec^* r \prec^* q \; , \; p \prec^* s \prec^* q \big\} \Big\vert \tau(p,q) = \tau_2 \Big\} \bigg)  d \theta  \end{equation}
On the other hand, if, instead of being a complex scalar, $\phi$ is assumed to be $n$-multiplet, then $a$ should be redefined as well: instead of being real-valued function it should become matrix-valued function. In other words, for any $r$ and $s$, the value of $a(r,s)$ is a hermitian $n \times n$ matrix. Once again, in order to define Lagrangian we will replace $\phi (z)$ with $e^{i a(x,z)} \phi (z)$, except that by exponential we mean matrix exponential. However, instead of substituting it into Eq \ref{ComplexLagrangian} we, instead, substitute it into Eq \ref{MultipletLagrangian} and obtain
\begin{equation}  {\mathcal L} (a,\phi; x) = \nonumber \end{equation}
\begin{equation} =  \frac{n}{\pi \tau_2^2 \mu (G)} \int_0^{2 \pi} \bigg(  \min \big\{( Re(e^{i \theta} \pi_1 U e^{i a(x,q)} \phi (q)) - Re(e^{i \theta} \pi_1 U e^{ia(x,p)} \phi (p)) )^2 \big\vert \nonumber \end{equation}
\begin{equation} \big\vert  p \prec x \prec q; \tau (p,q) = \tau_2 \big\} - \nonumber  \end{equation}
\begin{equation} - \min \Big\{ \max \big\{ ( ( Re(e^{i \theta} \pi_1 U e^{ia(x,r)} \phi (r)) - ( Re(e^{i \theta} \pi_1 U e^{ia(x,s)} \phi (s))  )^2 \big\vert \nonumber \end{equation}
\begin{equation} \big\vert  p \prec^* r \prec^* q \; , \; p \prec^* s \prec^* q \big\} \Big\vert \tau(p,q) = \tau_2 \Big\} \bigg)   \end{equation}

\subsection{Discretized circulation over $n$-point loop} \label{Loop}

So far we just done the coupling between scalar field and gauge field. Now we would like to move to the question of gauge field Lagrangian itself. As always, we go back to a continuum setting, with coordinates, where we do have $A_{\mu}$. In this setting, we take the Lagrangian density -- which is defined in terms of $A_{\mu}$ -- and re-express it in terms of $a$. After having done that, we take our result, written in terms of $a$, and simply copy it into the setting where we  no longer have coordinates.

In the first step, where we do coordinate-based calculation, the expression in terms of $A_{\mu}$ will be exact while the expression in terms of $a$ will be an approximation. On the other hand, in the discrete setting, without coordinates, the expression in terms of $a$ will become exact since it would be our new definition of the Lagrangian and the definitions are exact by their very nature. On the other hand, in the ``special case" of the manifold-like setting, the resulting expression involving $A_{\mu}$ will become an approximation. Since our construction is done in a local neighborhood, we can make an approximation where we assume the metric is Minkowskian, and the field is linear.

Since the approximation is the only thing we are interested in, we can drop approximation signs and simply assume in our calculation that the space is exactly Minkowskian and the electromagnetic field is exactly linear; and then we can drop that assumption and put approximation signs back once we get our result. Thus, throughout the following calculation, we will assume that
\begin{equation} g_{\mu \nu} = \eta_{\mu \nu} \end{equation}
\begin{equation} A_{\mu} (s) =A_{\mu} (r) + (s^{\nu} - r^{\nu}) \partial_{\nu} A_{\mu} \end{equation}
In light of Minkowskian setting, the geodesic connecting $r$ and $s$ will become a straight line. We can parametrize it as
\begin{equation} \gamma (\tau) = \frac{r+s}{2} + \tau \frac{s-r}{2} \; , \; -1 \leq \tau \leq 1 \end{equation}
Thus, $\gamma (-1) = r$ and $\gamma (1) =s$. If we assume $A_{\mu}$ varies linearly, we can write
\begin{equation} A_{\mu} (\gamma (\tau)) = A_{\mu} \bigg(\frac{r+s}{2} + \tau \frac{s-r}{2} \bigg) = A_{\mu} \bigg(\frac{r+s}{2} \bigg) + \tau \frac{s^{\nu} - r^{\nu}}{2} \partial_{\nu} A_{\mu} \end{equation}
If we substitute this into the integral
\begin{equation} a(r,s) = \int_{-1}^1 A_{\mu} \bigg(\frac{r+s}{2} + \tau \frac{s-r}{2} \bigg) d \bigg(\frac{r^{\mu}+s^{\mu}}{2} + \tau \frac{s^{\mu} -r^{\mu}}{2} \bigg) = \nonumber \end{equation}
\begin{equation} = \int_{-1}^1 \bigg(A_{\mu} \bigg(\frac{r+s}{2} \bigg) + \tau \frac{s^{\nu} - r^{\nu}}{2} \partial_{\nu} A_{\mu} \bigg) \frac{s^{\mu} -r^{\mu}}{2} d \tau = \nonumber \end{equation}
\begin{equation} = \frac{s^{\mu}-r^{\mu}}{2} A_{\mu} \bigg(\frac{r+s}{2} \bigg) \int_{-1}^1 d \tau + \frac{s^{\nu} - r^{\nu}}{2} \frac{s^{\mu} - r^{\mu}}{2} (\partial_{\nu} A_{\mu}) \int_{-1}^1 \tau d \tau = \nonumber \end{equation}
\begin{equation} = 2 \frac{s^{\mu} -r^{\mu}}{2} A_{\mu} \bigg(\frac{r+s}{2} \bigg)  = (s^{\mu} - r^{\mu}) A_{\mu} \bigg(\frac{r+s}{2} \bigg) = \nonumber \end{equation}
\begin{equation} = (s^{\mu} -r^{\mu}) \bigg(A_{\mu} (0) + \frac{r^{\nu} + s^{\nu}}{2} \partial_{\nu} A_{\mu} \bigg) = \nonumber \end{equation}
\begin{equation} = (s^{\mu} -r^{\mu})A_{\mu} (0) + \frac{(s^{\mu} -r^{\mu})(r^{\nu} +s^{\nu})}{2} \partial_{\nu} A_{\mu} = \nonumber \end{equation}
\begin{equation} = (s^{\mu} - r^{\mu})A_{\mu} (0) + \frac{s^{\mu}r^{\nu} + s^{\mu} s^{\nu} - r^{\mu} r^{\nu} - r^{\mu} s^{\nu}}{2} \partial_{\nu} A_{\mu} \label{a} \end{equation}
Now, consider the cycle, consisting of $n$ points. In order not to have to separately look at first and last point, we can number them by $k (mod \; n)$ where $cn+k (mod \; n) = k (mod \; n)$. Thus, our points would be $r_{0 (mod \; n)}, r_{1(mod \; n)}, \cdots, r_{n-1 (mod \; n)}$, and we are looking at the sum of the form $a(r_{0 (mod \; n)}, r_{1(mod \;n)}) + \cdots + a (r_{n-1 (mod \;n)}, r_{0 (mod \; n)})$. By using Eq \ref{a}, this becomes
\begin{equation} \sum_{k=0}^{n-1} ( a(r_{k  (mod \; n)},r_{k+1  (mod \; n)}) )=  \sum_{k=0}^{n-1} \bigg( (r_{k+1  (mod \; n)}^{\mu} - r_{k  (mod \; n)}^{\mu})A_{\mu} (0) + \end{equation}
\begin{equation} + \frac{r_{k+1  (mod \; n)}^{\mu}r_{k  (mod \; n)}^{\nu} + r_{k+1  (mod \; n)}^{\mu} r_{k+1  (mod \; n)}^{\nu} - r_{k  (mod \; n)}^{\mu} r_{k  (mod \; n)}^{\nu} - r_{k (mod \; n)}^{\mu} r_{k+1  (mod \; n) }^{\nu}}{2} \partial_{\nu} A_{\mu}  \bigg)  \nonumber \end{equation}
Observing that
\begin{equation} \sum_{k=0}^{n-1} (r_{k+1  (mod \; n)}^{\mu} - r_{k  (mod \; n)}^{\mu}) = 0 \end{equation}
\begin{equation}  \sum_{k=0}^{n-1} ( r_{k+1  (mod \; n)}^{\mu} r_{k+1  (mod \; n)}^{\nu} - r_{k  (mod \; n)}^{\mu} r_{k  (mod \; n)}^{\nu}) = 0 \end{equation}
it simplifies to
\begin{equation} \sum_{k=0}^{n-1} ( a(r_{k  (mod \; n)},r_{k+1  (mod \; n)}) )= \sum_{k=0}^{n-1} \frac{r_{k+1  (mod \; n)}^{\mu}r_{k  (mod \; n)}^{\nu}  - r_{k (mod \; n)}^{\mu} r_{k+1  (mod \; n) }^{\nu}}{2} \partial_{\nu} A_{\mu} \end{equation}
Observe that
\begin{equation} r_{k+1  (mod \; n)}^{\mu}r_{k  (mod \; n)}^{\nu} \partial_{\nu} A_{\mu}  = r_{k+1  (mod \; n)}^{\nu}r_{k  (mod \; n)}^{\mu} \partial_{\mu} A_{\nu} \end{equation}
to do the following calculation:  
\begin{equation} \sum_{k=0}^{n-1} a(r_{k  (mod \; n)},r_{k+1  (mod \; n)}) = \sum_{k=0}^{n-1} \frac{r_{k+1  (mod \; n)}^{\mu}r_{k  (mod \; n)}^{\nu}  - r_{k (mod \; n)}^{\mu} r_{k+1  (mod \; n) }^{\nu}}{2} \partial_{\nu} A_{\mu} = \nonumber \end{equation}
\begin{equation} = \sum_{k=0}^{n-1} \frac{r_{k+1  (mod \; n)}^{\mu}r_{k  (mod \; n)}^{\nu} \partial_{\nu} A_{\mu} - r_{k (mod \; n)}^{\mu} r_{k+1  (mod \; n) }^{\nu}\partial_{\nu} A_{\mu} }{2}  = \nonumber \end{equation}
\begin{equation} = \sum_{k=0}^{n-1} \frac{r_{k+1  (mod \; n)}^{\nu}r_{k  (mod \; n)}^{\mu} \partial_{\mu} A_{\nu} - r_{k (mod \; n)}^{\mu} r_{k+1  (mod \; n) }^{\nu}\partial_{\nu} A_{\mu} }{2}  = \nonumber \end{equation}
\begin{equation} = \sum_{k=0}^{n-1} \frac{r_{k  (mod \; n)}^{\mu} r_{k+1  (mod \; n)}^{\nu} \partial_{\mu} A_{\nu} - r_{k (mod \; n)}^{\mu} r_{k+1  (mod \; n) }^{\nu}\partial_{\nu} A_{\mu} }{2}  = \nonumber \end{equation}
\begin{equation} = \sum_{k=0}^{n-1} \frac{r_{k  (mod \; n)}^{\mu} r_{k+1  (mod \; n)}^{\nu} (\partial_{\mu} A_{\nu} - \partial_{\nu} A_{\mu}) }{2}  = \nonumber \end{equation}
\begin{equation} = \sum_{k=0}^{n-1} \frac{r_{k  (mod \; n)}^{\mu} r_{k+1  (mod \; n)}^{\nu}F_{\mu \nu}}{2}  = \nonumber \end{equation}
\begin{equation} = \frac{F_{\mu \nu}}{2}  \sum_{k=0}^{n-1} r_{k  (mod \; n)}^{\mu} r_{k+1  (mod \; n)}^{\nu} \label{CirculationGeneral} \end{equation}
We will now use this result for the two parts of Lagrangian, one called ``electric" and the other called ``magnetic", as described in Section \ref{Electric} and \ref{Magnetic} respectively. To avoid any confusion, these names are referring to electric and magnetic fields specifically in the reference frame of Alexandrov set; in more general reference frames they represent the mixture of the two, as described in Eq \ref{Einvariant} and Eq \ref{Binvariant}.

\subsection{Electric field in the reference frame of Alexandrov set} \label{Electric}

Suppose we have two fixed points $p \prec q$ and we are looking at all possible points $r \in \alpha (p,q)$ and $s \in \alpha (p,q)$. We would like to find the maxima and minima of
\begin{equation} \Phi = a(p,r) + a(r,q) + a(q,s) + a(s,p) \end{equation}
when $r$ and $s$ vary while $p$ and $q$ are fixed. Consider the reference frame in which $t$-axis passes through points $p$ and $q$, while electric field $\vec{E}$ points in $x^1$-direction. That is,
\begin{equation} p = \bigg(- \frac{\tau_2}{2}, 0, \cdots, 0 \bigg) \end{equation}
\begin{equation} q = \bigg( \frac{\tau_2}{2}, 0, \cdots, 0 \bigg) \end{equation}
\begin{equation} F_{0k} = - F_{k0} =  E \delta^1_k \end{equation}
From Eq \ref{CirculationGeneral} we have
\begin{equation} \Phi = a(p,r) + a(r,q) + a(q,s) + a (s,p) = \frac{1}{2} (F_{\mu \nu} p^{\mu} r^{\nu} + F_{\mu \nu} r^{\mu} q^{\nu} + F_{\mu \nu} q^{\mu} s^{\nu} + F_{\mu \nu} s^{\mu} p^{\nu}) = \nonumber \end{equation}
\begin{equation} = \frac{1}{2} \bigg(F_{01} p^0 r^1 + F_{10} r^1 q^0 + F_{01} q^0 s^1 + F_{10} s^1 p^0) = \nonumber \end{equation}
\begin{equation} = \frac{1}{2} \bigg(E \bigg(-\frac{\tau_2}{2} \bigg) r^1 + (-E) r^1 \frac{\tau_2}{2} + E \frac{\tau_2}{2} s^1 + (-E) s^1 \bigg(- \frac{\tau_2}{2} \bigg) \bigg) = \nonumber \end{equation}
\begin{equation} = \frac{1}{2} \bigg(- \frac{E \tau_2 r^1}{2} - \frac{E r^1 \tau_2}{2} + \frac{E \tau_2 S^1}{2} + \frac{E s^1 \tau_2}{2} \bigg) = \nonumber \end{equation}
\begin{equation} = \frac{1}{2} (-E \tau_2 r^1 + E \tau_2 s^1) = \frac{E \tau_2 (s^1-r^1)}{2} \end{equation}
Therefore, in order to maximize or minimize $\Phi$, we have to maximize or minimize $s^1 -r^1$. Given the constraint that $r$ and $s$ are within the Alexandrov set, it is easy to see that maximum is described as
\begin{equation} r_{max} = \bigg(0, - \frac{\tau_2}{2}, 0, \cdots, 0 \bigg) \end{equation}
\begin{equation} s_{max} = \bigg(0, \frac{\tau_2}{2}, 0, \cdots, 0 \bigg) \end{equation}
and minimum is described as
\begin{equation} r_{min} = \bigg(0, \frac{\tau_2}{2}, 0, \cdots, 0 \bigg) \end{equation}
\begin{equation} s_{min} = \bigg(0, - \frac{\tau_2}{2}, 0, \cdots, 0 \bigg) \end{equation}
Accordingly, the maximum and minimum values of $\Phi$ are given by
\begin{equation} \Phi_{max} = \frac{E \tau_2}{2} (s_{max}^1 -r_{max}^1) = \frac{E \tau_2}{2} \bigg(\frac{\tau_2}{2} - \bigg(- \frac{\tau_2}{2} \bigg) \bigg) = \frac{E \tau_2}{2} \tau_2 = \frac{E \tau_2^2}{2} \end{equation}
\begin{equation} \Phi_{min} = \frac{E \tau_2}{2} (s_{min}^1 -r_{min}^1) = \frac{E \tau_2}{2} \bigg(-\frac{\tau_2}{2} -  \frac{\tau_2}{2}  \bigg) =- \frac{E \tau_2}{2} \tau_2 = -\frac{E \tau_2^2}{2} \end{equation}
Since the minimum has negative sign, $\phi^2$ attains maximum there. Therefore, $\phi^2$ has two maximum points, with its common value being $E^2 \tau_2^4/4$. Thus,
\begin{equation} \max \{(a(p,r) + a(r,q) + a(q,s) + a(s,p))^2 \vert p \prec r \prec q \; , \; p \prec s \prec q \} = \frac{E^2 \tau_2^4}{4} \end{equation}
Note that
\begin{equation} E^2 \tau_2^4 = \tau_2^2 F_{0k} F_{0k} (q^0-p^0)(q^0-p^0)= \nonumber \end{equation}
\begin{equation} = - \tau_2^2 F_{0 \rho} F^{0 \rho} (q^0 -p^0)(q_0-p_0)= - \tau_2^2 F_{\mu \rho} F^{\nu \rho} (q^{\mu} - p^{\mu})(q_{\nu} - p_{\nu}) \end{equation}
So we can now go back to the general frame and write the covariant expression
\begin{equation} \max \{(a(p,r) + a(r,q) + a(q,s) + a(s,p))^2 \vert p \prec r \prec q \; , \; p \prec s \prec q \} = \nonumber \end{equation}
\begin{equation} = - \tau_2^2 F_{\mu \rho} F^{\nu \rho} (q^{\mu} - p^{\mu})(q_{\nu} - p_{\nu}) \end{equation}

\subsection{Magnetic field in the reference frame of Alexandrov set} \label{Magnetic}

Once again, lets consider the situation with $p \prec q$ and choose a coordinate system where $t$-axis passes through points $p$ and $q$. But, instead of looking at electric field, let us look at the magnetic field in that reference frame. The equator of Alexandrov set will simply become the intersection of $t=0$ hyperplane with its surface. Thus, a point $r$ is in equator if it is lightlike separated from points $p$ and $q$ at the same time. This is equivalent to saying that the Lorentzian distance from $p$ to $r$ as well as Lorentzian distance from $r$ to $q$ are both $0$. The discretized version of the statement that Lorentzian distance from $p$ to $r$ is $0$ is a statement that $p \prec r$ holds, but there is no point $s$ for which $p \prec s \prec r$ holds. This statement is denoted by $p \prec^* r$. Similarly, the statement that Lorentzian distance from $r$ to $q$ is $0$ is described as $r \prec^* q$. And, therefore, an equator of Alexandrov set consists of all points $r$ that obey $p \prec^* r \prec^* q$.

Let us consider $4$ points, $r_{0 (mod \; 4)}$, $r_{1 (mod \; 4)}$, $r_{2 (mod \; 4)}$ and $r_{3 (mod \; 4)}$ placed on equator of Alexandrov set; that is,
\begin{equation} \forall k \in \{0, 1, 2, 3 \} (p \prec^* r_{k (mod \; 4)} \prec^* q) \end{equation}
or, in coordinate-based language,
\begin{equation} \forall k \in \{0, 1, 2, 3 \} \bigg(r_{k (mod \; 4)}^0 =0 ; \vert \vec{r}_{k (mod \; 4)} \vert^2 = \frac{\tau_2^2}{4} \bigg) \end{equation}
From Eq \ref{CirculationGeneral} we know that
\begin{equation} \Phi = \sum_{k=0}^3 ( a(r_{k  (mod \; 4)},r_{k+1  (mod \; 4)}) ) = \frac{F_{\mu \nu}}{2}  \sum_{k=0}^3 r_{k  (mod \; 4)}^{\mu} r_{k+1  (mod \; 4)}^{\nu}  \label{0to3}\end{equation}
Note that
\begin{equation} \sum_{k=0}^3 ( a(r_{-k  (mod \; 4)},r_{-k-1  (mod \; 4)}) ) = -  \sum_{k=0}^3 ( a(r_{k  (mod \; 4)},r_{k+1  (mod \; 4)}) )  \end{equation}
and, therefore,
\begin{equation} \Phi_{min} = - \Phi_{max} \end{equation}
So our only task is to find $\Phi_{max}$. Lets do that by looking at variations:
\begin{equation} \frac{\partial \phi}{\partial r^{\rho}_{j (mod \; 4)}} = \frac{F_{\mu \nu}}{2}  \sum_{k=0}^3 (\delta^j_k \delta^{\mu}_{\rho} r_{k+1  (mod \; 4)}^{\nu} +  r_{k  (mod \; 4)}^{\mu} \delta^j_{k+1} \delta^{\nu}_{\rho}) = \nonumber \end{equation}
\begin{equation} =  \frac{F_{\mu \nu}}{2}  \bigg( \sum_{k=0}^3 \delta^j_k \delta^{\mu}_{\rho} r_{k+1  (mod \; 4)}^{\nu} + \sum_{k=0}^3  r_{k  (mod \; 4)}^{\mu} \delta^j_{k+1} \delta^{\nu}_{\rho} \bigg) =  \frac{F_{\mu \nu}}{2}  \bigg( \delta^{\mu}_{\rho} r_{j+1  (mod \; 4)}^{\nu} +   r_{j-1 (mod \; 4)}^{\mu} \delta^{\nu}_{\rho} \bigg) = \nonumber \end{equation}
\begin{equation} =  \frac{F_{\mu \nu}}{2} \delta^{\mu}_{\rho} r_{j+1  (mod \; 4)}^{\nu} +   \frac{F_{\mu \nu}}{2} r_{j-1 (mod \; 4)}^{\mu} \delta^{\nu}_{\rho} = \frac{F_{\rho \nu}}{2} r_{j+1  (mod \; 4)}^{\nu} +   \frac{F_{\mu \rho}}{2} r_{j-1 (mod \; 4)}^{\mu} = \nonumber \end{equation}  
\begin{equation} =  \frac{F_{\rho \sigma}}{2} r_{j+1  (mod \; 4)}^{\sigma} +   \frac{F_{\sigma \rho}}{2} r_{j-1 (mod \; 4)}^{\sigma} = \frac{F_{\rho \sigma}}{2} r_{j+1  (mod \; 4)}^{\sigma} -   \frac{F_{\rho \sigma}}{2} r_{j-1 (mod \; 4)}^{\sigma} = \nonumber \end{equation}
\begin{equation} =  \frac{F_{\rho \sigma}}{2} (r_{j+1  (mod \; 4)}^{\sigma} -  r_{j-1 (mod \; 4)}^{\sigma}) \label{SkipStep}  \end{equation}
Since we made an assumption that $r_{i (mod \; 4)}^0 =0$ for all $i$, we can replace the summation over spacetime indices $\rho$ and $\sigma$ with the summation over spacelike indices $r$ and $s$. So we obtain
\begin{equation}  \frac{\partial \phi}{\partial r^r_{j (mod \; 4)}} =\frac{F_{rs}}{2} (r_{j+1  (mod \; 4)}^s -  r_{j-1 (mod \; 4)}^s)  \label{rs} \end{equation}
Suppose $r_i$ are fixed for all $i \neq j$, and we would like to maximize $\Phi$ with respect to $r_{j (mod \; 4)}$. Note that we have a constraint
\begin{equation} \vert \vec{r}_{j(mod \; 4)} \vert^2 = \frac{\tau_2^2}{4} \end{equation}
Therefore, the maximum occurs if the gradient of $\Phi$ with respect to $\vec{r}_{j (mod \; 4)}$ is parallel to the gradient of $\vert \vec{r}_{j (mod \; 4)} \vert^2$ with respect to $\vec{r}_{j(mod \; 4)}$. But note that the gradient of $\vert \vec{r}_{j(mod \; 4)} \vert^2$ with respect to $\vec{r}_j$ is simply $2 \vec{r}_{j(mod \; 4)}$. Therefore, the gradient of $\Phi$ with respect to $\vec{r}_{j(mod \; 4)}$ should be parallel to $\vec{r}_{j(mod \; 4)}$ itself. Thus,
\begin{equation}  \frac{\partial \phi}{\partial r^r_{j (mod \; 4)}} = cr_j  \end{equation}
By substituting this into Eq \ref{rs}, we obtain
\begin{equation}  cr^r_{j (mod \; 4)}=  \frac{F_{rs}}{2} (r_{j+1  (mod \; 4)}^s -  r_{j-1 (mod \; 4)}^s) \label{cGrad} \end{equation}
We can find the value of $c$ from our knowldge that
\begin{equation} \vert \vec{r}_{j(mod \; 4)} \vert^2 = \frac{\tau_2^2}{4} \end{equation}
Thus,
\begin{equation} \frac{c^2 \tau_2^2}{4} = \frac{F_{rs_1}}{2} \frac{F_{rs_2}}{2} (r_{j+1  (mod \; 4)}^{s_1} -  r_{j-1 (mod \; 4)}^{s_1})(r_{j+1  (mod \; 4)}^{s_2} -  r_{j-1 (mod \; 4)}^{s_2}) \end{equation}
and, therefore,
\begin{equation} c = \frac{1}{\tau_2} \sqrt{F_{rs_1}F_{rs_2} (r_{j+1  (mod \; 4)}^{s_1} -  r_{j-1 (mod \; 4)}^{s_1})(r_{j+1  (mod \; 4)}^{s_2} -  r_{j-1 (mod \; 4)}^{s_2}) } \label{cOrig} \end{equation}
By going back to Eq \ref{cGrad} and replacing $j (mod \; 4)$ with $j+2 (mod \; 4)$ we obtain
\begin{equation} cr_{j+2 (mod \; 4)} = \frac{F_{rs}}{2} (r_{j+3  (mod \; 4)}^s -  r_{j+1 (mod \; 4)}^s) = \frac{F_{rs}}{2} (r_{j-1  (mod \; 4)}^s -  r_{j+1 (mod \; 4)}^s) = -cr_{j (mod \; 4)} \end{equation}
Thus,
\begin{equation} r_{j+2 (mod \; 4)} = - r_{j (mod \; 4)} \label{Minus} \end{equation}
By substituting this into Eq \ref{cOrig} we obtain
\begin{equation} c = \frac{1}{\tau_2} \sqrt{F_{rs_1}F_{rs_2} (r_{j+1  (mod \; 4)}^{s_1} -  r_{j-1 (mod \; 4)}^{s_1})(r_{j+1  (mod \; 4)}^{s_2} -  r_{j-1 (mod \; 4)}^{s_2}) } = \nonumber \end{equation}
\begin{equation}= \frac{1}{\tau_2} \sqrt{F_{rs_1}F_{rs_2} (r_{j+1  (mod \; 4)}^{s_1} +  r_{j+1 (mod \; 4)}^{s_1})(r_{j+1  (mod \; 4)}^{s_2} + r_{j+1 (mod \; 4)}^{s_2}) } = \nonumber \end{equation}
\begin{equation}= \frac{1}{\tau_2} \sqrt{F_{rs_1}F_{rs_2} (2 r_{j+1  (mod \; 4)}^{s_1} )(2r_{j+1  (mod \; 4)}^{s_2} ) } = \nonumber \end{equation}
\begin{equation}= \frac{2}{\tau_2} \sqrt{F_{rs_1}F_{rs_2} r_{j+1  (mod \; 4)}^{s_1} r_{j+1  (mod \; 4)}^{s_2} } \label{cSimple} \end{equation}
Also, by substituting Eq \ref{Minus} into Eq \ref{cGrad} we obtain
\begin{equation}  cr^r_{j (mod \; 4)}=  \frac{F_{rs}}{2} (r_{j+1  (mod \; 4)}^s -  r_{j-1 (mod \; 4)}^s) =   \frac{F_{rs}}{2} (-r_{j-1  (mod \; 4)}^s -  r_{j-1 (mod \; 4)}^s) = \nonumber \end{equation}
\begin{equation} =  \frac{F_{rs}}{2} (-2r_{j-1  (mod \; 4)}^s) = - F_{rs} r_{j-1  (mod \; 4)}^s \end{equation}
By using $j=k+1$ we obtain
\begin{equation}  cr^r_{k+1 (mod \; 4)}=- F_{rs} r_{k  (mod \; 4)}^s \label{SingleStep}\end{equation}
and, therefore
\begin{equation}  r^r_{k+1 (mod \; 4)}=- \frac{F_{rs}}{c} r_{k  (mod \; 4)}^s \end{equation}
By substituting this into Eq \ref{0to3} we obtain
\begin{equation} \Phi_{max}=  \frac{F_{\mu \nu}}{2}  \sum_{k=0}^3 r_{k  (mod \; 4)}^{\mu} r_{k+1  (mod \; 4)}^{\nu}  =  \frac{F_{rs}}{2}  \sum_{k=0}^3 r_{k  (mod \; 4)}^r r_{k+1  (mod \; 4)}^s= \nonumber \end{equation}
\begin{equation} = \frac{F_{rs}}{2}  \sum_{k=0}^3 r_{k  (mod \; 4)}^r  \bigg(- \frac{F_{st}}{c} r_{k  (mod \; 4)}^t \bigg)= \nonumber \end{equation}
\begin{equation} = - \frac{1}{2c} \sum_{k=0}^3 F_{rs} F_{ts} r_{k(mod \; 4)}^r r_{k (mod \; 4)}^t \end{equation}
From Eq \ref{cSimple} we know that, for any fixed $k$,
\begin{equation}  F_{rs} F_{ts} r_{k(mod \; 4)}^r r_{k (mod \; 4)}^t = \frac{c^2 \tau_2^2}{4} \end{equation}
and, therefore,
\begin{equation}  \sum_{k=0}^3 F_{rs} F_{ts} r_{k(mod \; 4)}^r r_{k (mod \; 4)}^t  = 4 \frac{c^2 \tau_2^2}{4} = c^2 \tau_2^2 \end{equation}
Thus, we obtain
\begin{equation} \Phi_{max} =- \frac{1}{2c} c^2 \tau_2^2 = - \frac{c \tau_2^2}{2} = \frac{\tau_2^2}{2}  \frac{2}{\tau_2} \sqrt{F_{rs_1}F_{rs_2} r_{j+1  (mod \; 4)}^{s_1} r_{j+1  (mod \; 4)}^{s_2} } = \nonumber \end{equation}
\begin{equation} = \tau_2  \sqrt{F_{rs_1}F_{rs_2} r_{j+1  (mod \; 4)}^{s_1} r_{j+1  (mod \; 4)}^{s_2} }= \tau_2  \sqrt{r_{j+1  (mod \; 4)}^{s_1} (-F_{s_1r}F_{rs_2}) r_{j+1  (mod \; 4)}^{s_2} } = \nonumber \end{equation}
\begin{equation} = \tau_2 \sqrt{r_{j+1  (mod \; 4)}^{s_1} (-F^2)_{s_1s_2}  r_{j+1  (mod \; 4)}^{s_2} } \end{equation}
This is true for all $j \in \{0, 1, 2, 3 \}$. This is not surprising since, due to Eq \ref{SingleStep}, the choice of location of any one of these $r_{j(mod \; 4)}$ dictates the locations of all the others. If we pick $j=3$, we obtain
\begin{equation} \Phi_{max} =\tau_2 \sqrt{r_{0 (mod \; 4)}^{s_1} (-F^2)_{s_1s_2}  r_{0  (mod \; 4)}^{s_2} } \end{equation}
We now have to determine the choice of $r_{0 (mod \;4)}$ that would maximize the above expression with respect to the constraint that its magnitude is $\frac{\tau_2}{2}$. Let $M$ be ``magnetic" matrix, that is to be obtained from removing the first row and first column from $F$:
\[ F = \left( \begin{array}{cccc}
0 & -E^1 & - E^2 & -E^3 \\
E^1 & 0 & B_3 & -B_2 \\
E^2 & -B_3 & 0 & B_1 \\
E^3 & B_2 & -B_1 & 0 \end{array} \right)\] \
\[ M = \left( \begin{array}{ccc}
0 & B_3 & -B_2 \\
-B_3 & 0 & B_1 \\
B_2 & -B_1 & 0 \end{array} \right)\] \
 Since the above expression doesn't involve any zero components, it can be rewritten as
\begin{equation} \Phi_{max} =\tau_2 \sqrt{r_{j+1  (mod \; 4)}^{s_1} (-M^2)_{s_1s_2}  r_{j+1  (mod \; 4)}^{s_2} } \end{equation}
In order to maximize it, we need to pick $r_{0(mod \; 4)}$ to be an eigenvector of $-M^2$ that corresponds to its largest possible eigenvalue. Let the set of eigenvalues of $-M^2$ be given by $\lambda_1 (-M^2), \cdots, \lambda_{d-1} (-M^2)$, and maximum eigenvalue be
\begin{equation} \lambda_{max} (-M^2) = \max (\lambda_1 (-M^2), \cdots, \lambda_{d-1} (-M^2)) \end{equation}
Thus we obtain
\begin{equation} \Phi_{max} = \frac{\tau_2^2}{2}  \sqrt{\max (\lambda_1 (-M^2), \cdots, \lambda_{d-1} (-M^2))} \end{equation}
 Therefore, the maximum will be proportional to the square root of the largest eigenvalue of $-F^2$, where $F$ is $d-1 \times d-1$ matrix with only spacelike components. In other words, we remove the row and column that corresponds to electric field, and only include the parts that correspond to magnetic field. Then, for $d=3$, we have
\[ M = \left( \begin{array}{ccc}
0 & B & 0 \\
-B & 0 & 0 \\
0 & 0 & 0 \end{array} \right)\] \
and, therefore
\begin{equation} -M^2 = -  \left( \begin{array}{ccc}
0 & B & 0 \\
-B & 0 & 0 \\
0 & 0 & 0 \end{array} \right) \left( \begin{array}{ccc}
0 & B & 0 \\
-B & 0 & 0 \\
0 & 0 & 0 \end{array} \right) = -  \left( \begin{array}{ccc}
-B^2 & 0 & 0 \\
0 & -B^2 & 0 \\
0 & 0 & 0 \end{array} \right) = \left( \begin{array}{ccc}
B^2 & 0 & 0 \\
0 & B^2 & 0 \\
0 & 0 & 0 \end{array} \right) \nonumber \end{equation}
So the largest eigenvalue is $B^2$ and its square root is $B$. So we obtain
\begin{equation} \Phi = \frac{B \tau_2^2}{2} \end{equation}

\subsection{The Lorentz invariance of the unwanted result} \label{Invariance}

In the above two subsections we have computed electric and magnetic field separately. On the first glance, this sounds like a violation of principles of relativity. Of course, after we found $\vec{E}$ and $\vec{B}$, we can manually compute $\vert \vec{E} \vert^2 - \vert \vec{B} \vert^2$. But the question is: why doesn't the Lorentz invariance arise naturally, given that causal structure is manifestly invariant?

In order to address this question, we first point out that the Alexandrov set we were working with \emph{does} have preferred reference frame: namely, its a reference frame in which $t$-axis passes through its poles, $p$ and $q$. Indeed, the same was true for scalar fields as well. In case of scalar fields, we got rid of the preferred frame by looking at all possible choices of $p$ and $q$ that obey certain constraints (namely, that their midpoint is $x$ and that the distance between them is $\tau_2$) and did the minimization. In other words, we were minimizing the maximum. For any given Alexandrov set, we were finding a maximum, and then we were doing minimization over all possible choices of Alexandrov set obeying our constraints. We propose to repeat the same thing here.

Now, in case of gauge field, the ``maxima" correspond to $\vert \vec{E} \vert^2$ and $\vert \vec{B} \vert^2$, while Alexandrov set corresponds to preferred frame. Thus ``minimizing the maximum" amounts to finding the reference frame in which $\vert \vec{E} \vert^2$ and $\vert \vec{B} \vert^2$ are the smallest. Now consider the following situation. Suppose Alice and Bob are moving with different velocities, and neither of them occupy the aforementioned reference frame. On the other hand, the reference frame that minimizes the value of $\vert \vec{E} \vert^2$ is occupied by Charlie. Now, if one asks Alice and Bob ``what is the value of $\vert \vec{E} \vert^2$ in your reference frame", they will give different answers. On the other hand, if one asks them ``what is the value of $\vert \vec{E} \vert^2$ in the reference frame where it is the smallest", they will both give the same answer: namely, they will both point to Charlie's reference frame and state the value of $\vert \vec{E} \vert^2$ that Charlie would observe. They will be able to do that without communicating to Charlie: their knowledge of $\vec{E}$ and $\vec{B}$ in \emph{their own} reference frame will enable them to determine the reference frame of Charlie, as well as the values of $\vec{E}$ and $\vec{B}$ that Charlie would see.  Importantly, in order to find out the value of $\vert \vec{E} \vert^2$ in Charlie's frame, the knowledge of $\vec{E}$ in their own frame is not enough: they also need to know $\vec{B}$. This means that they will actually be computing some \emph{function} of both $\vec{E}$ and $\vec{B}$ -- despite the fact that they would be answering a question about electric field alone (or magnetic field alone for that matter). That function would return the same value in both of their frames, while $\vert \vec{E} \vert^2$ would not.

This leads to the question: can that function be written in Lorentz invariant form? First of all, it can't possibly be $F^{\mu \nu} F_{\mu \nu}$ since the latter is $\vert \vec{E} \vert^2 - \vert \vec{B} \vert^2$ in Charlies frame, but we are looking for something that is $\vert \vec{E} \vert^2$ in Charlie's frame. What comes to our rescue is another Lorentz invariant expression, $\epsilon_{\mu \nu \rho \sigma} F^{\mu \nu} F^{\rho \sigma}$. I claim that both $\vert \vec{E} \vert^2$ in Charlies frame, as well as $\vert \vec{B} \vert^2$ in Charlie's frame are equal to some kind of linear combination of $F^{\mu \nu} F_{\mu \nu}$ and $\epsilon_{\mu \nu \rho \sigma} F^{\mu \nu} F^{\rho \sigma}$. Then, if one does the subtraction, $\vert \vec{E} \vert^2 -\vert \vec{B} \vert^2$, then $\epsilon_{\mu \nu \rho \sigma} F^{\mu \nu} F^{\rho \sigma}$ cancels out and we are left with $F^{\mu \nu} F_{\mu \nu}$. Now, we do not need it to cancel if our only goal is to have Lorentz invariance. The only purpose of cancelation is in order for us to match the experimentally verified theory.

Let us try to satisfy ourselves and explicitly show what kind of linear combinations we would get. I claim that the reference frame that minimizes $\vert \vec{E} \vert^2$ coincides with the reference frame that minimizes $\vert \vec{B} \vert^2$, and this happens to be the frame where $\vec{E}$ and $\vec{B}$ are parallel. So we need to show two things. First, we need to show that the reference frame where $\vec{E}$ and $\vec{B}$ are parallel exists. And then we have to show that this frame minimizes the values of $\vert \vec{E} \vert$ and $\vert \vec{B} \vert$.

Let us start from the reference frame where they are \emph{not} parallel. and find the way we should boost it so that they would be. We will pick a coordinate system in which both $\vec{E}$ and $\vec{B}$ are on $xy$-plane. That is,
\begin{equation} E_z = B_z = 0 \end{equation}
and we will move with velocity $v$ in $z$-direction relative to that frame. The Lorentz transformations of $\vec{E}$ and $\vec{B}$ are given by
\begin{equation} E_x^{\prime} = \frac{E_x - vB_y}{\sqrt{1-v^2}} \end{equation}
\begin{equation} E_y^{\prime} = \frac{E_y+vB_x}{\sqrt{1-v^2}} \end{equation}
\begin{equation} B_x^{\prime} = \frac{B_x+vE_y}{\sqrt{1-v^2}} \end{equation}
\begin{equation} B_y^{\prime} = \frac{B_y-vE_x}{\sqrt{1-v^2}} \end{equation}
Lets parametrize $v$ as
\begin{equation} v = \tanh \theta \end{equation}
Then we can rewrite the above Lorentz transformations as
\begin{equation} E_x^{\prime} = E_x \cosh \theta - B_y \sinh \theta \end{equation}
\begin{equation} E_y^{\prime} = E_y \cosh \theta + B_x \sinh \theta \end{equation}
\begin{equation} B_x^{\prime} = B_x \cosh \theta + E_y \sinh \theta \end{equation}
\begin{equation} B_y^{\prime} = B_y \cosh \theta - E_x \sinh \theta \end{equation}
\begin{equation} E_z^{\prime} = 0 \end{equation}
\begin{equation} B_z^{\prime} =0 \end{equation}
Now, $\vec{E}^{\prime}$ is parallel to $\vec{B}^{\prime}$ if and only if $E_x^{\prime} B_y^{\prime} - E_y^{\prime} B_x^{\prime} =0$. So let us compute $E_x^{\prime} B_y^{\prime} - E_y^{\prime} B_x^{\prime}$:
\begin{equation} E_x^{\prime} B_y^{\prime} - E_y^{\prime} B_x^{\prime} = \nonumber \end{equation}
\begin{equation} = ( E_x \cosh \theta - B_y \sinh \theta)(B_y \cosh \theta - E_x \sinh \theta) - ( E_y \cosh \theta + B_x \sinh \theta)(B_x \cosh \theta + E_y \sinh \theta) = \nonumber \end{equation}
\begin{equation} = E_xB_y \cosh^2 \theta - E_x^2 \cosh \theta \sinh \theta - B_y^2 \sinh \theta \cosh \theta + B_y E_x \sinh^2 \theta -  \nonumber \end{equation}
\begin{equation} - E_y B_x \cosh^2 \theta - E_y^2 \cosh \theta \sinh \theta - B_x^2 \sinh \theta \cosh \theta - B_x E_y \sinh^2 \theta = \nonumber \end{equation}
\begin{equation} = (E_x B_y-E_yB_x) \cosh^2 \theta  + (B_y E_x - B_x E_y) \sinh^2 \theta - (B_x^2 +B_y^2 + E_x^2 +E_y^2) \sinh \theta \cosh \theta = \nonumber \end{equation}
\begin{equation} = (E_x B_y-E_yB_x)(\cosh^2 \theta + \sinh^2 \theta) -  (B_x^2 +B_y^2 + E_x^2 +E_y^2) \sinh \theta \cosh \theta = \nonumber \end{equation}
\begin{equation} = (E_x B_y-E_yB_x) \cosh 2 \theta -  (B_x^2 +B_y^2 + E_x^2 +E_y^2) \frac{\sinh 2 \theta}{2}  \label{EtimesB}\end{equation}
Therefore, they are parallel when
\begin{equation} E_x^{\prime} B_y^{\prime} - E_y^{\prime} B_x^{\prime} =0 \Longleftrightarrow \tanh 2 \theta = \frac{2(E_xB_y -E_y B_x)}{B_x^2 +B_y^2 +E_x^2+E_y^2} \label{EquationWithTanh}\end{equation}
So we would like to know whether or not there is $\theta$ that satisfies the above. Observe that
\begin{equation} (E_x-B_y)^2 \geq 0 \Longrightarrow E_x^2 - 2E_xB_y + B_y^2 \geq 0 \Longrightarrow  2E_xB_y  \leq E_x^2 +B_y^2 \end{equation}
\begin{equation} (E_x+B_y)^2 \geq 0 \Longrightarrow E_x^2 +2E_xB_y + B_y^2 \geq 0 \Longrightarrow  2E_xB_y  \geq  - E_x^2 -B_y^2 \end{equation}
\begin{equation} (E_y -B_x)^2 \geq 0 \Longrightarrow E_y^2 - 2E_yB_x + B_x^2 \geq 0 \Longrightarrow 2 E_yB_x  \leq E_y^2 +B_x^2 \end{equation}
\begin{equation} (E_y +B_x)^2 \geq 0 \Longrightarrow E_y^2 + 2E_yB_x + B_x^2 \geq 0 \Longrightarrow 2 E_yB_x  \geq -E_y^2 -B_x^2 \end{equation}
This implies that we have
\begin{equation} \vert 2E_xB_y \vert \leq E_x^2 +B_y^2 \end{equation}
\begin{equation} \vert  2E_yB_x \vert  \leq   E_y^2 +B_x^2 \end{equation}
Therefore, by triangle inequality,
\begin{equation} \vert 2 (E_xB_y -E_y B_x) \vert \leq \vert 2E_x B_y \vert + \vert 2E_y B_x \vert \leq E_x^2 +B_y^2 + E_y^2 +B_x^2 \end{equation}
and, therefore,
\begin{equation} \bigg\vert \frac{2(E_xB_y -E_y B_x)}{B_x^2 +B_y^2 +E_x^2+E_y^2}  \bigg\vert \leq 1 \label{Leq} \end{equation}
Now, in order for Eq \ref{EquationWithTanh} to have a solution we need to be able to replace $\leq$ sign with $<$ sign in Eq \ref{Leq}. This can be done if \emph{either} we replace $(E_x-B_y)^2 \geq 0$ with $(E_x-B_y)^2 > 0$, or we replace $(E_y -B_x)^2 \geq 0$ with $(E_y -B_x)^2 > 0$, or both. This, in turn, would happen if either $E_x \neq \pm B_y$ or $E_y \neq \pm B_x$ or both. If we have $+$ in place of $\pm$, then we can obtain $\neq$ by making slight rotatoin if we happened to be in the frame with $=$. But if we have $-$ in place of $\pm$ then it might be impossible in case that the two vectors happened to be orthogonal with an equal magnitude. Thus, we need to rule out the latter situation. This can be ruled out on statistical basis. Since causal set is countable, we have a zero probability of hitting a certain point in an uncountable set. Based on this argument we replace $\leq$ with $<$ and conclude that Eq \ref{EquationWithTanh} always has a solution. This means that there is, indeed, a frame where $\vec{E}$ and $\vec{B}$ are parallel.

Let us now show that that frame minimizes both $\vert \vec{E} \vert^2$ and $\vert \vec{B} \vert^2$. The value of $\vert \vec{E} \vert^2$ is given by the following calculation:
\begin{equation} (E_x^{\prime})^2 + (E_y^{\prime})^2 = (E_x \cosh \theta - B_y \sinh \theta)^2 + ( E_y \cosh \theta + B_x \sinh \theta)^2 = \nonumber \end{equation}
\begin{equation} = E_x^2 \cosh^2 \theta - 2E_x B_y \cosh \theta \sinh \theta + B_y^2 \sinh^2 \theta + E_y^2 \cosh^2 \theta + 2E_y B_x \cosh \theta \sinh \theta + B_x^2 \sinh^2 \theta = \nonumber \end{equation}
\begin{equation} =  (E_x^2 + E_y^2) \cosh^2 \theta + (B_x^2 + B_y^2) \sinh^2 \theta - 2 \sinh \theta \cosh \theta (E_xB_y - E_y B_x)  \end{equation}
On the other hand, the value of $\vert \vec{B} \vert^2$ is computted as
\begin{equation} (B_x^{\prime})^2 + (B_y^{\prime})^2 = (B_x \cosh \theta + E_y \sinh \theta)^2 + ( B_y \cosh \theta - E_x \sinh \theta)^2 = \nonumber \end{equation}
\begin{equation} = B_x^2 \cosh^2 \theta + 2B_x E_y \cosh \theta \sinh \theta + E_y^2 \sinh^2 \theta + \nonumber \end{equation}
\begin{equation} +  B_y^2 \cosh^2 \theta - 2B_y E_x \cosh \theta \sinh \theta + E_x^2 \sinh^2 \theta = \nonumber \end{equation}
\begin{equation} = (B_x^2 +B_y^2) \cosh^2 \theta + (E_x^2 +E_y^2) \sinh^2 \theta - 2 \sinh \theta \cosh \theta (E_x B_y - E_y B_x) \end{equation}
Observe that
\begin{equation} \cosh^2 \theta + \sinh^2 \theta = \cosh 2 \theta \end{equation}
\begin{equation} \cosh^2 \theta - \sinh^2 \theta =1 \end{equation}
and, therefore,
\begin{equation} \cosh^2 \theta = \frac{1 + \cosh 2 \theta}{2} \end{equation}
\begin{equation} \sinh^2 \theta = \frac{\cosh 2 \theta -1}{2} \end{equation}
By making these substitutions, we obtain
\begin{equation} (E_x^{\prime})^2 +(E_y^{\prime})^2 =  (E_x^2 + E_y^2) \cosh^2 \theta + (B_x^2 + B_y^2) \sinh^2 \theta - 2 \sinh \theta \cosh \theta (E_xB_y - E_y B_x)  = \nonumber \end{equation}
\begin{equation} =   (E_x^2 + E_y^2)\frac{1 + \cosh 2 \theta}{2}+ (B_x^2 + B_y^2) \frac{\cosh 2 \theta -1}{2}- 2 \sinh \theta \cosh \theta (E_xB_y - E_y B_x)  = \nonumber \end{equation}
\begin{equation} = \frac{E_x^2 + E_y^2 - B_x^2 - B_y^2}{2} + \frac{E_x^2 +E_y^2 +B_x^2+B_y^2}{2} \cosh 2 \theta - (E_xB_y-E_yB_x) \sinh 2 \theta \end{equation}
as well as
\begin{equation} (B_x^{\prime})^2 + (B_y^{\prime})^2 =  (B_x^2 +B_y^2) \cosh^2 \theta + (E_x^2 +E_y^2) \sinh^2 \theta - 2 \sinh \theta \cosh \theta (E_x B_y - E_y B_x) = \nonumber \end{equation}
\begin{equation} = (B_x^2 +B_y^2)\frac{1 + \cosh 2 \theta}{2}  + (E_x^2 +E_y^2)  \frac{\cosh 2 \theta -1}{2}  - \sinh 2 \theta (E_x B_y - E_y B_x) = \nonumber \end{equation}
\begin{equation} = \frac{B_x^2+B_y^2-E_x^2-E_y^2}{2} + \frac{E_x^2 +E_y^2 +B_x^2+B_y^2}{2} \cosh 2 \theta - (E_xB_y-E_yB_x) \sinh 2 \theta \end{equation}
Thus, we see that $(E_x^{\prime})^2 +(E_y^{\prime})^2$ differs from $(B_x^{\prime})^2 + (B_y^{\prime})^2$ by a quantity that is constant in $\theta$. That explains why the frame that minimizes one of them would minimize the other. The derivative of the above is given by
\begin{equation} \frac{d}{d \theta} ((E_x^{\prime})^2 +(E_y^{\prime})^2) = \frac{d}{d \theta} ((B_x^{\prime})^2 + (B_y^{\prime})^2) = \nonumber \end{equation}
\begin{equation} = (E_x^2 +E_y^2 +B_x^2+B_y^2) \sinh 2 \theta - 2 (E_xB_y-E_yB_x) \cosh 2 \theta \end{equation}
By comparing this to Eq \ref{EtimesB} we have
\begin{equation} \frac{d}{d \theta} ((E_x^{\prime})^2 +(E_y^{\prime})^2) = \frac{d}{d \theta} ((B_x^{\prime})^2 + (B_y^{\prime})^2) =  -2 (E_x^{\prime} B_y^{\prime} - E_y^{\prime} B_x^{\prime}) \end{equation}
Thus, the derivative is $0$ if and only if $\vec{E}$ nad $\vec{B}$ are parallel.

Let us now return back to the Alice and Bob problem we talked about earlier. We have just shown that, in Charlies reference frame, $\vec{E}$ and $\vec{B}$ are parallel. In that frame, we can do the following calculations:
\bea
& &E^2 - B^2 = F^{\mu\nu}\, F_{\mu\nu} \\
& &EB = \frac{1}{k}\, \epsilon_{\alpha\beta\gamma\delta}\,
F^{\alpha\beta}\, F^{\gamma \delta}\;.
\eea
This gives us a system of two equations and two unknowns, which means that if we solve it for $E^2$ and $B^2$, we will get a covariant expression for each of these terms separately. This should not surprise us. After all, when we decided to treat $E$ and $B$ as scalars, we were assuming that we are in a special frame where $E$ and $B$ are parallel. Thus, the procedure of {\em first\/} finding such frame and {\em then\/} evaluating $E$ and $B$ in that frame {\em is\/} covariant, despite the fact that $E$ and $B$ in an arbitrary frame are {\em not}.

Now, to satisfy ourselves, let us solve that system of equations to get an expression for $E$ and $B$. The second equation implies that $B = \epsilon_{\alpha\beta\gamma\delta}\, F^{\alpha\beta}\, F^{\gamma\delta}/kE$. Substituting this into the first equation, we get
\beq
F^{\mu\nu}\, F_{\mu\nu} = E^2 - \frac{(\epsilon_{\alpha\beta\gamma\delta}\,
F^{\alpha\beta}\, F^{\gamma\delta})^2}{k^2\, E^2}\;.
\eeq
Multiplying it by $E^2$, and moving all terms to the left-hand side, we get the following equation
\beq
E^4 - E^2\, F^{\mu\nu}\, F_{\mu\nu} - \frac{(\epsilon_{\alpha\beta\gamma\delta}\,
F^{\alpha\beta}\, F^{\gamma\delta})^2}{k^2\, E^2} = 0\;.
\eeq
This solves to
\beq
E^2 = \frac12\,\bigg[F^{\mu\nu}\, F_{\mu\nu} + \sqrt{(F^{\mu\nu}\, F_{\mu\nu})^2 + \frac{4}{k^2}\, (\epsilon_{\alpha\beta\gamma\delta}\, F^{\alpha\beta}\, F^{\gamma\delta})^2} \bigg]\;.
\label{Einvariant} \eeq
Thus, first equation of the system of two equations gives us
\beq
B^2 = \frac12\,\bigg[F^{\mu\nu}\, F_{\mu\nu} + \sqrt{(F^{\mu\nu}\, F_{\mu\nu})^2 + \frac{4}{k^2}\, (\epsilon_{\alpha\beta\gamma\delta}\, F^{\alpha\beta}\, F^{\gamma\delta})^2} \bigg] - F^{\mu\nu}\,F_{\mu\nu}\;.
\label{Binvariant} \eeq
As we see, these are definitely covariant expressions. The reason we are not getting the answer we would like is that there is more than one covariant contraction: $F^{\mu \nu}\, F_{\mu\nu}$ and $\epsilon_{\alpha \beta \gamma \delta}\, F^{\alpha \beta}\, F^{\gamma \delta}$. We would like to only have the former and get rid of the latter. That is what we will do in Section \ref{ElectromagneticResult}.

\subsection{Full electromagnetic Lagrangian} \label{ElectromagneticResult}

Let us now summarize what we have found by writing the expressions for the full electromagnetic Lagrangian. Just like we did with scalar field, the Lagrangian has two parts, ${\mathcal L}_E$ and ${\mathcal L}_B$, while the full Lagrangian is the difference between the two. These two parts are defined as follows:
\begin{equation} {\mathcal L}_E (a; \alpha(x,y)) = \frac{1}{\tau_2^2} \min \Big\{ \max \big\{ (a (p,r) + a(r,q) + a(q,s) + a(s,p))^2 \big\vert p \prec r \prec q ;p \prec s \prec q \big\} \Big\vert \nonumber \end{equation}
\begin{equation} \Big\vert p \prec x \prec q \; , \; \tau (p,q) = \tau_2 \Big\} \end{equation}
\begin{equation} {\mathcal L}_B(a; \alpha(x,y))= \frac{1}{\tau_2^2} \min \bigg\{ \max \bigg\{ \bigg(\sum_{k=1}^k a(r_{k (mod \; 4)}, r_{k+1 (mod \; 4)}) \bigg)^2 \bigg\vert \forall k (p \prec^* r_{k (mod \; 4)}  \prec^* q) \bigg\} \bigg\vert \nonumber \end{equation}
\begin{equation} \bigg\vert p \prec x \prec q \; , \; \tau (p,q) = \tau_2 \bigg\} \end{equation}
\begin{equation} {\mathcal L} (a; x) = {\mathcal L}_E (a; \alpha(x,y)) -  {\mathcal L}_B(a; \alpha(x,y)) \end{equation}
which, upon substitution, becomes
\begin{equation} {\mathcal L} (a,x) =  \frac{1}{\tau_2^2} \bigg( \min \Big\{ \max \big\{ (a (p,r) + a(r,q) + a(q,s) + a(s,p))^2 \big\vert p \prec r \prec q ;p \prec s \prec q \big\} \Big\vert \nonumber \end{equation}
\begin{equation} \Big\vert p \prec x \prec q \; , \; \tau (p,q) = \tau_2 \Big\} - \label{EMfull} \end{equation}
\begin{equation} -   \min \bigg\{ \max \bigg\{ \bigg(\sum_{k=1}^k a(r_{k (mod \; 4)}, r_{k+1 (mod \; 4)}) \bigg)^2 \bigg\vert \forall k (p \prec^* r_{k (mod \; 4)}  \prec^* q) \bigg\} \bigg\vert \nonumber \end{equation}
\begin{equation} \bigg\vert p \prec x \prec q, \tau (p,q) = \tau_2 \bigg) \nonumber \end{equation}
In coordinate-based setting when the fields behave linearly, these result in
\begin{equation} {\mathcal L}_E (a; \alpha(x,y)) \approx  \frac12\,\bigg[F^{\mu\nu}\, F_{\mu\nu} + \sqrt{(F^{\mu\nu}\, F_{\mu\nu})^2 + \frac{4}{k^2}\, (\epsilon_{\alpha\beta\gamma\delta}\, F^{\alpha\beta}\, F^{\gamma\delta})^2} \bigg] \end{equation}
\begin{equation} {\mathcal L}_B (a; \alpha(x,y)) \approx \frac12\,\bigg[F^{\mu\nu}\, F_{\mu\nu} + \sqrt{(F^{\mu\nu}\, F_{\mu\nu})^2 + \frac{4}{k^2}\, (\epsilon_{\alpha\beta\gamma\delta}\, F^{\alpha\beta}\, F^{\gamma\delta})^2} \bigg] - F^{\mu\nu}\,F_{\mu\nu} \end{equation}
\begin{equation} {\mathcal L} (a; \alpha(x,y)) \approx F^{\mu\nu}\,F_{\mu\nu} \end{equation}

\subsection{Non-abelian gauge theory}

The group $SU(n)$ is generated by $n^2-1$ Hermitian matrices, $T_1, \cdots, T_{n^2-1}$. Let us denote $F^i_{\mu \nu} T_i$ by $F_{\mu \nu}$:
\begin{equation} F_{\mu \nu} = F^i_{\mu \nu} T_i \end{equation}
and, therefore,
\begin{equation} F^i_{\mu \nu} = \frac{1}{2} Tr (F_{\mu \nu} T_i) \end{equation}
We can then perform the following calculation:
\begin{equation} F_i^{\mu \nu} F_{i \mu \nu} = \frac{1}{2} Tr (F^{\mu \nu} F_{\mu \nu}) = \frac{1}{2 \mu (G)} \int Tr ((UF)^{\mu \nu} (UF)_{\mu \nu}) d \mu (U) =  \end{equation}
\begin{equation} = \frac{1}{2\mu (G)} \int \sum_{k=1}^{n^2-1} Tr ((T_k UF)^{\mu \nu} (T_k UF)_{\mu \nu}) d \mu (U) = \nonumber \end{equation}
\begin{equation} = \frac{1}{2\mu (G)} \int \sum_{k=1}^{n^2-1} Tr ((T_1 UF)^{\mu \nu} (T_1 UF)_{\mu \nu}) d \mu (U) = \nonumber \end{equation}
\begin{equation} = \frac{1}{2\mu (G)} \bigg(\sum_{k=1}^{n^2-1}  1 \bigg) \int Tr ((T_1 UF)^{\mu \nu} (T_1 UF)_{\mu \nu}) d \mu (U) = \nonumber \end{equation}
\begin{equation} = \frac{n^2-1}{2\mu (G)}\int Tr ((T_1 UF)^{\mu \nu} (T_1 UF)_{\mu \nu}) d \mu (U) \nonumber \end{equation}
Therefore, we will re-interpret $a(r,s)$ as a Hermitian matrix, replace $a(r,s)$ with $T_1 U a(r,s)$, and then take a trace and integrate. If we start with Eq \ref{EMfull} and then to the above, we obtain
\begin{equation} {\mathcal L} (a,x) =  \frac{1}{2 \mu (G)\tau_2^2} \bigg[ \min \Big\{ \max \big\{ (Tr (T_1 U(a (p,r) + a(r,q) + a(q,s) + a(s,p))))^2 \big\vert \nonumber \end{equation}
\begin{equation} \big\vert p \prec r \prec q ;p \prec s \prec q \big\}  \Big\vert p \prec x \prec q \; , \; \tau (p,q) = \tau_2 \Big\} -  \end{equation}
\begin{equation} -  \min \bigg\{ \max \bigg\{ \bigg(Tr \bigg(T_1 U \sum_{k=1}^k a(r_{k (mod \; 4)}, r_{k+1 (mod \; 4)}) \bigg) \bigg)^2 \bigg\vert \nonumber \end{equation}
\begin{equation} \forall k (p \prec^* r_{k (mod \; 4)}  \prec^* q) \bigg\} \bigg\vert  \tau(p,q) = \tau_2 \bigg\} \bigg] \nonumber \end{equation}

\section{Conclusions}

In this paper, we have accomplished two things. First of all, we have expanded the discussion of QFT on causal sets from just scalar field to gauge field. Secondly, we have found a way of treating both of these fields in such a way that avoids explicit dimensional dependence in the ratio between the coefficients. The dimensional dependence of those ratios in other approaches made it seem like relativistic invariance was ``forced". On the other hand, in the current paper, it appeared in much more natural way. We still had to artificially insert the subtractions (see Eq \ref{ScalarSubtraction} and Eq \ref{EMfull}), but we were still able to avoid the insertion of non-trivial coefficients there. Furthermore, even without these subtractions, the results would still be Lorentz invariant, just not the ones that we like. For scalar field, they are given by Eq \ref{LtScalar} and Eq \ref{LsScalar}. For electromagnetic field, they are given by Eq \ref{Einvariant} and Eq \ref{Binvariant}. But then the subtraction of these equations from one another would lead to the cancelation of unwanted terms, and produce the results we are looking for

Apart from the conceptual advantage, the fact that there are no dimension-dependent coefficients would potentially allow for the proposed model of the scalar field to be adopted into the theories where dimensionality changes in small scales \cite{2Dim1, 2Dim2}, or disappears altogether. In case of gauge field it is a bit more problematic since we assumed $3+1$ dimensions in some of our discussion.

One weakness in the current model, however, is the fact that maximization and minimization is nonlinear and nonquadratic. This makes it virtually impossible to compute path integrals analytically -- although one can still attempt to compute them numerically. It is important to emphasize that path integrals still exists as a mathematical objects and, therefore, one still expects an interference and so forth. But those things can not be computted analytically. This, in itself, does not falsify the theory as such. For example, the lack of analytic solutions to Einstein equation does not falsify the Einstein equation as such. At the same time, however, the fact that the options that \emph{do} appear quadratic exist (such as \cite{Delambertian1, Delambertian2, paper1, paper3, edges} among others) is a good reason to take the current model with a grain of salt. But, as we mentioned previously, these other models have non-trivial dimension-dependent coefficients, while the current model does not. So, in light of these pros and cons, we believe that all these models -- along with the one proposed in the current paper -- should be put side by side. This will enable the researchers to make their own decision as to which to utilize based on their specific line of research.

There was a further work done by the current author where higher order nonlinear corrections to the above model were computed \cite{NonlinearCorrections}. This might potentially provide an empirical way of testing the theory. However it should be noted that this calculation is based on the assumption that all the functions are smooth and well behaved, which is not true when it comes to Feynmann path integral. So its predictions are classical rather than quantum mechanical. At the same time, in light of the fact that the size of Alexandrov set is very small, it is hard to think of a situation when this would be relevant for classical calculations. Therefore, as it stands, the results of \cite{NonlinearCorrections} are really just mathematical exercises as opposed to actual physical predictions. Nevertheless, it might still provide an illustration of the kind of nonlinear effects we might be looking for in our future research of quantum version of this.

It is interesting to note that there is other work that investigates non-linearities in the context of quantum physics (see \cite{QuantumCorrections}). Since those nonlinearities are unrelated to causal sets, they likely take completely different form from the ones that we have. However, it is interesting to investigate whether, by some chance, there are some similarities. And, if so, what would be underlying reason for them. We are leaving it for the future research.

{\bf Declaration:}

Funding: I am currently a graduate student at the University of New Mexico and, as such, I am a Teaching Assistant, so I get paid for teaching.

Conflicts of interest/Competing interests: While the current paper is single author, the Proceeding (cited below) was joint work with Luca Bombelli. However, seeing that the Proceedings below is a lot shorter than the current paper, and the presentation is different, I regard it as a separate paper, so I don't see any conflict of interest. 

Availability of data and material: It is available on arXiv:0807.4709. Also, one of the older versions of this was posted on Proceedings for ``Workshop on Continuous and Lattice Approaches to Quantum Gravity 17-19 September 2008" However, the Proceeding paper is a lot shorter: it is only 8 pages. So it is abbreviated in a lot of ways. For example, it doesn't do a lot of the explicit calculations that are being done here. It also focuses exclusively on real scalar field and electromagnetic field, but it doesn't do complex scalar fields, spin 0 multiplets and non-commutative gauge fields. 

Code availability: I used latex. While I submitted the PDF file, I still have the TeX version of it at the computer at home. 

Author contribution: The current paper is single author, although the Proceeding (cited above) was joint work with Luca Bombelli.


\begin{thebibliography}{77}

\bibitem{Nowhere} Christian Wüthrich, Nick Huggett ``Out of Nowhere: Spacetime from causality: causal set theory" arXiv:2005.10873

\bibitem{Surya} Sumati Surya, ``Causal set approach to quantum gravity" Living Reviews in Relativity volume 22, Article number: 5 (2019)  and arXiv:1903.11544

\bibitem{Hawking} S W Hawking, A R King and P J McCarthy 1976 “A new topology for curved spacetime

which incorporates the causal, differential and conformal structures” J. Math. Phys. 17 174-181.

\bibitem{Malament} D Malament 1977 “The class of continuous timelike curves determines the topology of

spacetime” J. Math. Phys. 18 1399-1404.

\bibitem{Solodukhin} G. W. Gibbons, S. N. Solodukhin ``The Geometry of Small Causal Diamonds" Phys.Lett.B649:317-324,2007 and  arXiv:hep-th/0703098

\bibitem{Distance1} G. Brightwell and R. Gregory, 1991, “Structure of random discrete spacetime” Phys. Rev. Lett. 66, 260–263 (1991)

\bibitem{Distance2} R. Ilie, G.B. Thompson, D.D. Reid 2006 “A numerical study of the correspondence between paths in a causal set and geodesics in the continuum” Class. Quantum Grav. 23 3275, and arXiv:gr-qc/0512073.

\bibitem{Delambertian1} Fay Dowker, Lisa Glaser ``Causal set d'Alembertians for various dimensions" 2013, Class. Quantum Grav. 30 195016 and arXiv:1305.2588

\bibitem{Delambertian2} Siavash Aslanbeigi, Mehdi Saravani, Rafael D. Sorkin ``Generalized Causal Set d'Alembertians" JHEP 1406 (2014) 024 DOI 10.1007/JHEP06(2014)024 and arXiv:1403.1622

\bibitem{2Dim1} S. Carlip ``Dimension and Dimensional Reduction in Quantum Gravity" arXiv:1904.04379

\bibitem{2Dim2} Joshua H. Cooperman, Manuchehr Dorghabekov ``Setting the physical scale of dimensional reduction in causal dynamical triangulations" Phys. Rev. D 100, 026014 (2019) and arXiv:1812.09331

\bibitem{paper1} R Sverdlov and L Bombelli 2008 ``Gravity and matter in causal set theory''\hfill\break {\tt arXiv:0801.0240v2}.

\bibitem{paper3} R Sverdlov 2008 ``Gauge fields in causal set theory'' {\tt arXiv:0807.2066}.

\bibitem{edges} R. Sverdlov ``Electromagnetic Lagrangian on a causal set that resides on edges rather than points" arXiv:1805.08064

\bibitem{Nonlocality} Sorkin, D. P. (20 March 2007). "Does Locality Fail at Intermediate Length-Scales". arXiv:gr-qc/0703099

\bibitem{NonlinearCorrections} R. Sverdlov ``Non-linear corrections to Lagrangians predicted by causal set theory: Flat space bosonic toy model" arXiv:1201.5850

\bibitem{QuantumCorrections} T.F.Kamalov, Quantum Correction for Newton's Law of Motion, Symmetry 2020, 12, 63

\bibitem{Chains} Joachim Kambor and Nomaan X ``Manifold Properties from Causal Sets using Chains" arXiv:2007.03835

\bibitem{Haar1} Jonathan Gleason, ``Existence and uniqueness of Haar measure" https://meduza.io/feature/2018/09/22/kazhdaya-storona-osoznaet-sebya-zhertvoy-eto-sostoyanie-blizkoe-k-paranoye

\bibitem{Haar2} Luis J. Boya, E.C.G. Sudarshan, Todd Tilma ``Volumes of Compact Manifolds" https:math-ph/0210033.pdf

\bibitem{Percolation} D. P. Rideout, R. D. Sorkin ``A Classical Sequential Growth Dynamics for Causal Sets" Phys.Rev.D61:024002,2000 and  arXiv:gr-qc/9904062

\bibitem{Robb1} A.A. Robb, The absolute relations of time and space, (Cambridge U. P. , Cambridge, 1921. (80p)).

\bibitem{Robb2} A. A. Robb, Geometry of Time and Space (Cambridge University, Cambridge, England, 1936).

\bibitem{GaugeFields} Roman Sverdlov ``Gauge Fields in Causal Set Theory" arXiv:0807.2066

\bibitem{EMedges} Roman Sverdlov ``Electromagnetic Lagrangian on a causal set that resides on edges rather than points" arXiv:1805.08064

\bibitem{ContinuousMeasurement} Roman Sverdlov ``Continuous measurement on a causal set with and without a boundary" arXiv:2006.08724

\bibitem{PhaseSpacetime} Roman Sverdlov ``Causal set as a discretized phase spacetime" arXiv:0910.2498

\bibitem{Proceedings} Roman Sverdlov ``Quantum field theory on a causal set", PoS LATTICE2018 (2019) 234.

\bibitem{ManifoldlikeThesis} Miremad Aghili, ``Manifoldlike causal sets" Ph.D. Thesis, University of Mississippi. https://egrove.olemiss.edu/etd/1529/

\bibitem{Dynamics} David Rideout, ``Dynamics of Causal Sets", Ph.D. Thesis https://cds.cern.ch/record/597403/files/0212064.pdf 

\bibitem{DiscreteGravity} Rafael D. Sorkin, ``Causal Sets: Discrete Gravity" https://cds.cern.ch/record/640720/files/0309009.pdf



\end{thebibliography}
\end{document}